\documentclass{aa} % for a referee version
\usepackage{graphics}

\def\lsim{~\rlap{$<$}{\lower 1.0ex\hbox{$\sim$}}}
\def\gsim{~\rlap{$>$}{\lower 1.0ex\hbox{$\sim$}}}

\begin{document}

   \thesaurus{20    % A&AS Section 20: General
              (11.09.1 \object{Mrk~86};  % Galaxies: individual
               11.09.5;  % Galaxies: irregular,
               11.19.4;  % Galaxies: star clusters,
               11.19.5)} % Galaxies: stellar content.

   \title{Mapping the star formation history of \object{Mrk~86}$^*$}
   \subtitle{I. Data and models}

   \author{A. Gil de Paz\inst{1}\and J. Zamorano\inst{1}\and
   J. Gallego\inst{1}\and F. de B. Dom\'{\i}nguez\inst{2}}

   \offprints{A. Gil de Paz\\$^*$Visiting Astronomer (AGdP), Kitt
   Peak National Observatory, National Optical Astronomy
   Observatories, which is operated by the Association of Universities
   for Research in Astronomy, Inc. (AURA) under cooperative agreement
   with the National Science Foundation.}

   \institute{Astrof\'\i sica y CC. de la Atm\'osfera,
   Universidad Complutense de Madrid, Avda. Complutense s/n, E-28040
   Madrid, Spain\and OTRI, Universidad Carlos III de Madrid, C/ Butarque, 15,
   E-28911 Legan\'es, Spain\\ e-mail: gil@astrax.fis.ucm.es (AGdP)}

   \date{Received May 19 2000; accepted June 8, 2000}

   \maketitle

   \label{firstpage}

   \begin{abstract}

We have obtained optical ($BVR$, [\ion{O}{iii}]$\lambda$5007\AA\ and
H$\alpha$), near infrared ($JHK$) imaging and long-slit optical
spectroscopy for the Blue Compact Dwarf galaxy \object{Mrk~86}
(\object{NGC~2537}). In this paper, the first of two, we present
optical-near-infrared colors and emission-line fluxes for the
currently star-forming regions, intemediate aged starburst and
underlying stellar population. We also describe the evolutionary
synthesis models used in Paper~II.

The $R$ and H$\alpha$ luminosity distributions of the galaxy
star-forming regions show maxima at $M_R$=$-$9.5$^{\mathrm{m}}$ and
$L_{\mathrm{H}\alpha}$=10$^{37.3}$\,erg\,s$^{-1}$. The underlying
stellar population shows an exponential surface brigthness profile
with central value, $\mu_{\mathrm{E,0}}$=21.5\,mag\,arcsec$^{-2}$, and
scale, $\alpha$=0.88\,kpc, both measured in the $R$-band image. In the
galaxy outer regions, dominated by this component, no significant
color gradients are observed.

Finally, a complete set of evolutionary synthesis models have been
developed, covering a wide range in metallicity,
1/50\,Z$_{\sun}$$<$Z$<$2\,Z$_{\sun}$, and burst strength,
1-10$^{-4}$. These models include nebular continuum and recombination
and forbidden-line emission.


\end{abstract}

\section{Introduction}
\label{introduction}

The high star formation rate (SFR hereafter) and low neutral gas
content deduced for dwarf star-forming galaxies imply consumption
time-scales of about 10$^{9}$\,yr (Fanelli et al. \cite{fanelli};
Thuan \& Martin \cite{thuan81}), much shorter than the age of the
Universe. Searle et al. (\cite{searle73}) suggested that either these
objects are truly young systems or they have an intermittent star
formation history with short intense star-forming episodes followed by
long quiescent phases. Although this question remained unanswered
during decades (Thuan \cite{thuan83}; Campbell \& Terlevich
\cite{campbell}; Loose \& Thuan \cite{loose85}), nowadays, most of the
studies on dwarf star-forming systems, including Blue Compact Dwarf
galaxies (BCD hereafter), have revealed the existence of an evolved
underlying population (Kunth et al. \cite{kunth88}; Hoffman et
al. \cite{hoffman}; Papaderos et al. \cite{papaII}; Doublier
\cite{doublier98}; Norton \& Salzer \cite{norton}).

Therefore, the understanding of the mechanism, or mechanisms,
governing this star formation regulation is the stepping stone of the
dwarf star-forming galaxies evolution. Moreover, this mechanism
constitutes the missing link between the different dwarf star-forming
galaxies (Silk et al. \cite{silk87}; Burkert \cite{burkert};
Drinkwater \& Hardy \cite{drinkwater}; Papaderos et
al. \cite{papaII}).

There are two different approaches to address these questions. On the
one hand, statistical analysis of a sample of these objects will allow
the study of the relationships between fundamental parameters and
properties of the star-forming dwarf galaxies (Marlowe et
al. \cite{marlowe95}; Papaderos et al. \cite{papaI},
\cite{papaII}). On the other hand, the detailed analysis of individual
low-redshift objects with multiple regions of star formation is
fundamental to reconstruct their star formation histories. In
particular, we can obtain a better understanding of possible effects
of merging or some internal process as self-propagation that could
originate the spreading of star formation, and the effects of these
star-forming events on the future star formation.

Among dwarf galaxies, Blue Compact Dwarf galaxies conform the
subpopulation where the star-forming events are most violent. Blue
Compact Dwarfs are low-luminosity (M$_B$
$\ge$$-$18$^{\mathrm{m}}$)\footnote{For
$H_0$=75\,km\,s$^{-1}$\,Mpc$^{-1}$ (Thuan \& Martin
\cite{thuan81}).} galaxies with compact sizes whose spectra are
similar to those of low metallicity \ion{H}{ii} regions (Searle \&
Sargent \cite{searle72}; Kunth \& Sargent \cite{kunth86}; Thuan \&
Martin \cite{thuan81}). Their spectra are characterized by emission
lines over a blue continuum which implies the existence of a large
fraction of OB stars and an intense star-forming activity.

The Blue Compact Galaxy \object{Mrk~86}=\object{NGC~2537} (Shapley \&
Ames \cite{shapley32}; Markarian \cite{markarian69}), also known as
Arp~6 (Arp \cite{arp66}), constitutes an excellent laboratory to test the BCD
star formation history since its star-forming regions populate all the
galaxy. This object is the propotype of the iE galaxies, the most
important BCD galaxies subclass, that conform 70 per cent of the BCD
galaxies (Thuan \& Martin \cite{thuan81}; Thuan \cite{thuan91}).

Optical imaging (both broad- and narrow-band) and near infrared images
have been obtained to map the different regions and get insight into
the star formation history of every region. In addition, we have
obtained very high quality spectra with spatial resolution. The
spectra of 22 star-forming knots and of free interim were taken with
10 distinct slit positions and orientations and moderate and high
spectral resolution.

After briefly introduce the previous results on \object{Mrk~86} in
Sect.~\ref{mrk86}, we will describe the observations and data
reduction in Sect.~\ref{observations}. The data analysis is presented
in Sect.~\ref{analysis}. We describe the evolutionary synthesis models
in Sect.~\ref{models}. In Sect.~\ref{results} we give some
results. Finally, our conclusions are presented in
Sect.~\ref{conclusions}.

In Gil de Paz et al. (\cite{paperII}; Paper~II hereafter) we derive
the physical properties of the star-forming regions and stellar
populations of \object{Mrk~86} comparing our photometric data with the
evolutionary synthesis models presented in this paper. Electron
densities and temperatures and chemical abundances of the ionized gas
in the galaxy star-forming regions will be also obtained. Finally, a
global interpretation of the galaxy will be given.

\setcounter{table}{0}
\begin{table}
\begin{centering} \caption{Collection of the published parameters of \object{Mrk~86}}
\begin{tabular}{llc}
Parameter & Data & Reference\\[10pt]
\hline
v$_{\sun}$ & 447\,km\,s$^{-1}$ & (1)\\
v$_{\mathrm{LG}}$ & 522\,km\,s$^{-1}$ & (1)\\
v$_{\mathrm{LG}}$ & 460\,km\,s$^{-1}$ & (2)\\
Distance          & 6.9\,Mpc      & (3) \\
$B$ & 12.8 & (4)\\
{\cal f}(1510\AA) & 0.87$\times$10$^{-14}$\,erg\,m$^{-2}$\,s$^{-1}$\,\AA$^{-1}$ & (5)\\
S(2.8cm) & 7$\pm$2\,mJy & (6)\\
S(6.3cm) & 18$\pm$4\,mJy & (7)\\
S(1.2cm) & 11$\pm$4\,mJy & (7)\\
{\cal f}(12$\mu$m) & 0.25\,Jy & (8)\\
{\cal f}(25$\mu$m) & 0.42\,Jy & (8)\\
{\cal f}(60$\mu$m) & 3.15\,Jy & (8)\\
{\cal f}(100$\mu$m) & 6.26\,Jy & (8)\\
$L_{\mathrm{IR}}$ & 0.35\,10$^{9}$\,L$_{\sun}$ & (8)\\
$J$ & 13.25 & (9)\\
$H$ & 12.57 & (9)\\
$K$ & 12.37 & (9)\\
$L$ & 12.63 & (9)\\
M(\ion{H}{i}) & 2.4\,10$^8$\,M$_{\sun}$ & (10)\\
M$_T$ & 5.8\,10$^8$\,M$_{\sun}$ & (10)\\
{\cal f}(\ion{H}{i} 21cm) & 2.1\,Jy\,km\,s$^{-1}$ & (11)\\
FWHM$_{\mathrm{\ion{H}{i}}}$ & 93\,km\,s$^{-1}$ & (11)\\
I$_{\mathrm{CO}}$ & 0.83$\pm$0.12\,K\,km\,s$^{-1}$    & (12)\\
\hline
\end{tabular}
\medskip
\end{centering}
\\(1) Bottinelli et al. (\cite{bottinelli90}); (2) Rood \& Dickel (\cite{rood76}); 
(3) Sharina et al. (\cite{sharina}); (4) Dultzin-Hacyan et
al. (\cite{dultzin90}); (5) Fanelli et al. (\cite{fanelli}); (6) Klein
et al. (\cite{klein91}); (7) Klein et al. (\cite{klein84}); (8)
Thronson \& Telesco (\cite{thronson86}); (9) Thuan (\cite{thuan83},
measured with 7\farcs8 apertures); (10) Thuan \& Martin
(\cite{thuan81}); (11) Verter (\cite{verter}); (12) Sage et
al. (\cite{sage92}).
\label{biblio}
\end{table}

\section{Mrk 86}
\label{mrk86}

\object{Mrk~86} ($\alpha$(2000)=08$^{h}$13$^{m}$14.56$^{s}$ 
$\delta$(2000)=$+$45\degr 59\arcmin 30.2\arcsec)\footnote{Center of
the outer isophotes as measured in the $R$-band (see
Sect.~\ref{analysis:sbp})} is a well-known object among BCD galaxies
(Loose \& Thuan \cite{loose85}; Sage et al. \cite{sage92}; see
Table~\ref{biblio}). In the Loose \& Thuan (\cite{loose85}) BCD
classification, \object{Mrk~86} belongs to the iE class: smooth
elliptical Low Surface Brightness (LSB) underlying stellar component
on which several knots of star formation are superimposed. The present
star formation activity is spread out in several clumps over a
relatively large fraction of its entire surface. In fact,
\object{Mrk~86} shows the highest ratio of the area covered by
star-forming regions to the total projected area of the galaxy within
the sample of Papaderos et al. (\cite{papaI}).

To this moment \object{Mrk~86} has not been the subject of a single
and proof study, inasmuch the large number of published data on it. A
collection of the published data is given in Table~\ref{biblio}. A
more detailed description of the previous observations is given by Gil
de Paz et al. (\cite{gil1}; GZG hereafter).

\subsection{Distance}
\label{distance}

\object{Mrk~86} is a nearby star-forming galaxy. The heliocentric recession 
velocity is $v_{\sun}$=447\,km\,s$^{-1}$ (see recession velocities
referred to the Local Group in Table~\ref{biblio}).

Using the mean blue magnitude of the three brightest blue stars
Sharina et al. (\cite{sharina}) derived a distance for this object of
6.9\,Mpc with an uncertainty of 20~per cent (M. E. Sharina, private
communication). We can also estimate the distance to \object{Mrk~86}
if we assume that it is gravitationally bounded to the edge-on spiral
galaxy \object{UGC~4278} (see Paper~II). \object{UGC~4278} has a
corrected $I$-band absolute magnitude of $-$18.21$\pm$0.11 (see
Giovanelli et al. \cite{giovanelli97}) with a logarithmic 21\,cm line
width of 2.220$\pm$0.019 (in km\,s$^{-1}$). Therefore, with respect to
the Giovanelli et al. (\cite{giovanelli97}) Tully-Fisher template,
this galaxy shows an offset of 0.66$^{\mathrm{m}}$, which corresponds
to a peculiar radial velocity of $-$253\,km\,s$^{-1}$. Thus, its
recesion velocity in the CMB reference frame should be
968\,km\,s$^{-1}$ ($v_{\sun}$=565\,km\,s$^{-1}$), which implies a
minimum distance for \object{Mrk~86} of 13\,Mpc (for
$H_{0}$=75\,km\,s$^{-1}$\,Mpc$^{-1}$ and assuming that the difference
in heliocentric velocity is due to orbital motion). The large
difference between this value and that given by Sharina et
al. (\cite{sharina}) suggests that these objects are probably not
physically bounded. Thus, since no other distance indicator is
available for \object{Mrk~86}, we have used the distance measured by
Sharina et al. (\cite{sharina}) throughout this work. This value leads
a projected spatial scale of 33\,pc\,arcsec$^{-1}$.

\section[]{Observations and reduction}
\label{observations}

\subsection{Optical imaging}
\label{optimage}

An optical $B$-band image was obtained at the 1-m Jacobus Kapteyn
Telescope (JKT) of Roque de los Muchachos observatory (La Palma,
Spain) in 1997 November with a 24\,$\mu$m 1024$\times$1024 pixels
Tek\#4 CCD (see Table~\ref{journal}). An additional Johnson-$V$ image
was taken at the 1.52-m spanish telescope at EOCA (Calar Alto,
Almer\'{\i}a, Spain) in December 1993 with a Tek 1024$\times$1024 CCD
with pixel size of 19\,$\mu$m. This observation was broken up into
five exposures with a total integration time of 2400\,s. Finally, an
$R$-band image was obtained during service time in November 1998 with
the ING wide-field camera (ING-WFC) equipped with four EEV42
2k$\times$4k pixels CCD detectors on the 2.5-m Isaac Newton Telescope
(La Palma, Spain).

Narrow-band images in the light of [\ion{O}{iii}]$\lambda$5007\AA\
($\lambda_{0}$ = 5012\AA , FWHM = 50\AA) and H$\alpha$ ($\lambda_{0}$
= 6568\AA , FWHM = 95\AA) were obtained. In order to subtract the
continuum, $V$ and $R$-band images were used, respectively. The
[\ion{O}{iii}]$\lambda$5007\AA\ image was secured for us on December
1993 during service time with a 1280$\times$1180 pixels EEV5 CCD
attached to the 2.5-m Isaac Newton Telescope. The H$\alpha$ image was
taken during the same service time observations that the $R$-band
image in November 1998 using the ING-WFC camera at the 2.5-m Isaac
Newton Telescope (La Palma, Spain).

High S/N ratio dome flats and exposures of the sky taken in twilight
were obtained in any case. The standard procedure of bias removal,
dark-current subtraction, and flat fielding using dome and sky
flat-field images was performed using the ESO image processing system
MIDAS for the $V$ image and IRAF\footnote{IRAF is distributed by the
National Optical Astronomy Observatories, which are operated by the
Association of Universities for Research in Astronomy, Inc., under
cooperative agreement with the National Science Foundation.} for the
$B$ and $R$-band images.

Atmospheric conditions were photometric during the observing runs. $B$
and $V$ images were flux calibrated observing repeatedly during the
nights a set of standard stars, taken from the lists given by Kent
(\cite{kent}) and Landolt (\cite{landolt}). Finally, the $R$-band
image was flux calibrated using the radial surface brightness profile
published by Papaderos et al. (\cite{papaI}).

The [\ion{O}{iii}]$\lambda$5007\AA\ and H$\alpha$ images flux
calibration was performed as follows (see also GZG). The continuum
emission was subtracted, using the $V$ and $R$ images,
respectively. Then, we trimmed those regions covered by the slits of
the $b$(blue) and $r$(red) spectra (see Table~\ref{journal}). The $b$
and $r$ spectra were convolved with the transmission curves of the
corresponding narrow-band filters. Then, we compared image counts and
fluxes measured in the convolved spectra for several adjacent regions
in order to determine the precise location of the slits. Finally, once
the regions covered by the slits were precisely defined, we corrected
these calibration relations for the sensitivity of the filter at the
corresponding wavelength and, in the case of the H$\alpha$ flux, for
the contribution of the [\ion{N}{ii}]$\lambda$6548\AA\ and
[\ion{N}{ii}]$\lambda$6583\AA\ lines. The discrepancies obtained among
different spectra were in all cases lower than 10 per cent.

\begin{table}
\caption[]{Journal of observations.}
\begin{tabular}{lrlcc}
& \multicolumn{4}{c}{Spectroscopic Observations$^{\dagger}$}\\
          & Exp. & Slit & Range &   Disp.   \\ 
Telescope & time(s)  &  & (nm)  & (\AA/pix) \\
\hline
{CAHA2.2m} & {3600} & \#1,2,4,6$b$   & 330-580 & {2.6} \\
{CAHA2.2m} & {1800} & \#5$b$         & 330-580 & {2.6} \\
{CAHA2.2m} & {3600} & \#1,2,4,5,6$r$ & 435-704 & {2.6} \\
{CAHA2.2m} & {3600} & \#3            & 390-650 & {2.6} \\
{INT2.5m}  & {1800} & \#7,8R         & 637-677 & {0.39} \\ 
{INT2.5m}  & {900}  & \#9R           & 637-677 & {0.39} \\ 
\hline
&    \multicolumn{4}{c}{Image Observations}\\
          & Exp. & Filter & Scale            & PSF  \\
Telescope & time(s) &        & (\arcsec /pixel) & (\arcsec) \\
\hline
{JKT1.0m}  &   {600} & $B$       & 0\farcs33  & 1\farcs0 \\ 
{CAHA1.5m} &  {2400} & $V$       & 0\farcs33  & 1\farcs6 \\ 
{INT2.5m}  &   {900} & $R$       & 0\farcs333 & 1\farcs2 \\ 
{INT2.5m}  &   {900} & [\ion{O}{iii}]$\lambda$5007 & 0\farcs57 & 2\farcs5 \\ 
{INT2.5m}  &  {7200} & H$\alpha$ & 0\farcs333 & 1\farcs2 \\ 
{KPNO2.3m} &   {900} & $J$       & 0\farcs66 & 1\farcs8 \\
{KPNO2.3m} &   {360} & $H$       & 0\farcs66 & 1\farcs6 \\
{KPNO2.3m} &   {540} & $K$       & 0\farcs66 & 1\farcs7 \\
\hline
\end{tabular}
$^{\dagger}$ See Fig.~1 of GZG for slit orientations and positions.
\label{journal}
\end{table}

\begin{figure*}
%\resizebox{\hsize}{!}{\includegraphics{fig1.ps}} 
\vspace{18.5cm}
\caption{$BVRJHK$, H$\alpha$ and [\ion{O}{iii}]$\lambda$5007\AA\ images of \object{Mrk~86}.}
\label{images}
\end{figure*}

\begin{figure*}
\begin{center}
%\resizebox{14.5cm}{!}{\includegraphics{fig2.ps}} 
\vspace{14.5cm}
\end{center}
\caption{Knot contours derived using the {\sc cobra} program superimposed on the H$\alpha$ image. The $thick$-$lined$ contour corresponds to the starburst component as described by Papaderos et al. (\cite{papaI}, see Sect.~\ref{central}). The crossed-circle position marks the center of the outer $R$-band isophotes.}
\label{myconts}
\end{figure*}

\subsection{Near-infrared imaging}
\label{nirimage}

Near-infrared images (nIR hereafter) of \object{Mrk~86} in $J$
($\lambda _{0}$ = 1.25$\mu$m, FWHM = 0.30$\mu$m), $H$ ($\lambda _{0}$
= 1.65$\mu$m, FWHM = 0.28$\mu$m) and $Ks$ ($\lambda _{0}$ =
2.15$\mu$m, FWHM = 0.33$\mu$m) were obtained on January 1998 with the
Steward Observatory near-infrared camera equipped with a
256$\times$256 {\sc nicmos3} detector attached to the 2.3-m Bok
Telescope at Kitt Peak National Observatory (Arizona, USA). The
observational procedure closely follows that of Arag\'on-Salamanca et
al. (\cite{aas93}) and Gil de Paz et al. (\cite{nirucm}). The total
integration time of each subimage was broken up into
background-limited sub-exposures to avoid saturation of the
detector. Images of adjacent blank areas of the sky were alternated
with the target frames for accurate flat-fielding. Comparable amounts
of time were spent imaging the source and the sky to ensure adequate
monitoring of the sky changes. Individual subexposures of the object
were offset several arcseconds to improve the final result.

The reduction process include: (1) bias and dark subtraction of object
and sky frames; (2) flat-fielding using normalized sky frames; (3) sky
subtraction using sky frames taken before and after each exposure; (4)
bad pixel removal; (5) registering of the subimages using fractional
pixel shifts and (6) median combining of all individual frames. The
reduction was carried out using own IRAF procedures.

\begin{figure}
\resizebox{\hsize}{!}{\includegraphics{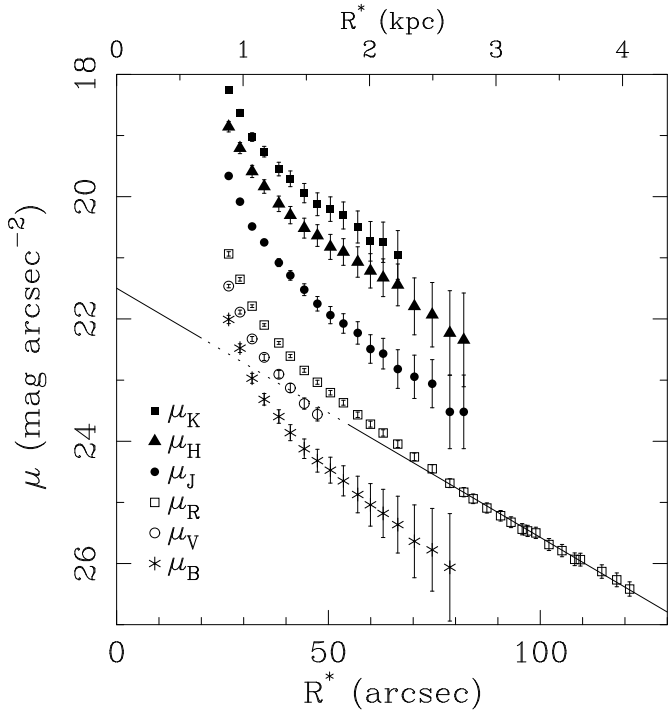}} 
\caption{Surface brightness profiles in the $BVRJHK$ bands. The fit 
to the exponential component of the $R$-band profile at distances
larger than 70\arcsec\ is also drawn. Error bars represent
$\pm$1$\sigma$. The $K$-band profile has been offset $-$0.5$^{\mathrm{m}}$ in order to avoid confusion with the $H$-band profile.}
\label{profiles}
\end{figure}

The nIR images were calibrated observing standard stars from the list
from Elias et al. (\cite{elias}) during the nights at the same
airmasses than the object. We have assumed a color independent
correction between the $Ks$ and $K'$ bands ($K' \equiv KM$; $\lambda
_{0}$ = 2.12$\mu$m, FWHM = 0.34$\mu$m; Wainscoat \& Cowie
\cite{wainscoat}). In order to check the validity of this assumption, we 
convolved the $Ks$ and $K'$ filters response functions with Planck
spectral distributions at different temperatures in the range
3000-20000\,K, obtaining the fluxes $F_{Ks}$ and $F_{K'}$. The larger
difference in 2.5$\times$log($F_{Ks}$/$F_{K'}$) within this
temperature range was 0.025$^{\mathrm{m}}$, small enough to assume
this correction to be independent of the spectral energy distribution.

Thus, once the fluxes were transformed to $K'$-band fluxes, we
converted them to the standard $K$-band ($\lambda _{0}$ = 2.19$\mu$m,
FWHM = 0.41$\mu$m) applying the standard correction given by Wainscoat \&
Cowie (\cite{wainscoat}), $K'-K$=0.22$\times$($H-K$).

In Fig.~\ref{images} we show the \object{Mrk~86} neighborhood in the
optical-near-infrared bands studied, including the H$\alpha$ and
[\ion{O}{iii}]$\lambda$5007\AA\ images. The astrometric calibration
was performed using the program {\sc PlateAstrom} (Garc\'{\i}a-Dab\'o
\& Gallego \cite{dabo}; see also {\tt
http://www.ucm.es/info/Astrof/opera/opera.html}). The bright field
star placed at the relative position (60\arcsec E,30\arcsec N),
saturated in the $V$ and $R$-band images, was artificially removed.

\subsection{Spectroscopy}
\label{optspectra}

Up to 14 optical long-slit spectra at 10 different slit positions were
obtained. The 11 medium-low resolution spectra ($b$ and $r$ spectra;
see Table~\ref{journal}) were obtained with the Boller \& Chivens
spectrograph at the Cassegrain focus of the 2.2-m telescope at the
German-Spanish Calar Alto observatory (Almer\'{\i}a, Spain) in January
1993. Althought special care was taken to place the slits into
position, using offsets from the bright field star at north of the
galaxy, there were some lack of precision. The actual positions of the
slits (see Fig.~1 of GZG) were determined {\sl a posteriori}
comparing broad-band image spatial cuts with the spatial profiles of
the spectra previously convolved with the corresponding filter
transmission curves. The detector employed was a 1024$\times$1024
Tek\#6 CCD with a pixel size of 24\,$\mu$m. The 600\,gr\,mm$^{-1}$
grating chosen provided a spectral resolution of 6\AA\ in the light of
H$\alpha$ and a reciprocal dispersion of 2.6\AA/pixel, with a slit
width of 2$\farcs$65. The final spatial scale was 1$\farcs$43/pixel.
The spectral coverage was around 2500\AA\ and the grating angle was
selected to cover the blue region ($\sim$3300-5800\AA) and red domain
($\sim$4350-7045\AA) in two different exposures which overlapped on
the H$\beta$ region. The seeing was variable during the observing run
with FWHM between 1$\farcs$5--2$\farcs$5. The low airmasses at which
these spectra were obtained ($\leq$1.2) guarantee that no significant
loss of blue light due to atmospheric refraction has ocurred. In
addition, three high resolution spectra (7R,8R,9R) were obtained with
the IDS spectrograph at the Isaac Newton Telescope (INT) of the Roque
de los Muchachos Observatory (La Palma, Spain) in January 1998 with a
1024$\times$1024 Tek\#3 24\,$\mu$m CCD (see Table~\ref{journal}). The
1200R grating (1200\,gr\,mm$^{-1}$) was used with an slit width of
1\arcsec, providing a spectral resolution of 0.9\AA\ in the light of
H$\alpha$ and a reciprocal dispersion of 0.39\AA/pixel. The spatial
scale was of 0.33\arcsec/pixel.

These spectra were reduced using the {\sc figaro} (January 1993) and
IRAF (January 1998) software packages. After bias removal and
flat-fielding, the frames were cleaned of cosmic rays. The sky was
removed from each frame by subtracting a polynomial fitted to those
regions free for object emission. Wavelength calibration was performed
by using He-Ar lamps observed inmediately before and after the galaxy
integration. The standard stars Hiltner~102 and Hiltner~600 were
observed at different airmasses in order to correct for atmospheric
extinction and to ensure absolute flux calibration.

Finally, we requested an UV spectrum of \object{Mrk~86} from the
International Ultraviolet Explorer (IUE) Final Archive (see
Fig.~\ref{uvspectra}). It was originally taken by Alloin and Duflot
in January 1983 (see Bonatto et al. \cite{bonatto}). The total
exposure time of this spectrum, SWP18927, was 24000\,s. It was
obtained in low dispersion mode with the SWP camera. In the observing
spectral range, 1150-1975\AA, the resolution power varies between
270-300 (Cassatella et al. \cite{cassatella}).

\section{Analysis}
\label{analysis}

\subsection{Surface brightness profiles}
\label{analysis:sbp}

We have analyzed the surface brightness profile of \object{Mrk~86} in
different bands ($BVRJHK$; see Fig.~\ref{profiles}). These surface
brightness profiles have been obtained using the IRAF task {\sc
ellipse}. Background galaxies and foreground stars were interactively
masked. We obtained the surface brightness profiles over the original
images, excepting in the outer regions of the $R$-band profile where a
median filter of 5$\times$5 pixels (0\farcs333/pixel) was applied. Due
to the contribution of about 60 high surface brightness regions, we
only fitted the isophotes with major axis larger than 25\arcsec. Using
the mean of the isophotal centers measured between 80 and 120\arcsec\
in the $R$-band image we have estimated the galaxy coordinates given
in Sect.~\ref{mrk86} (see also Fig.~\ref{myconts}).

\subsection{Knot positions and sizes}
\label{pos+sizes}
 
\begin{table}
\caption[]{Positions and sizes for the \object{Mrk~86} neighborhood regions.}
\begin{tabular}{|lccccc|}
\hline
\# \ \ \#' & RA(2000) & DEC(2000) & $r_{1e}$ \ $r_{1e}$' & $e$ & Cl. \\
\hline
(1) \ (2)& (3) & (4) & (5)\ \ (6) &(7)&(8)\\
\hline
01     \,\ \ \ -- &  08:13:16.08 & $+$46:00:17.1& 0.71  PLR      & 2 & B \\
02     \,\ \ \ -- &  08:13:16.33 & $+$46:00:12.1& 0.85  0.45     & 1 & B \\
03     \,\ \ \ -- &  08:13:15.91 & $+$46:00:11.8& 0.74  0.17     & 2 & B \\
04     \,\ \   01 &  08:13:14.72 & $+$46:00:04.1& 1.13  0.87     & 2 & F \\
05     \,\ \ \ -- &  08:13:13.38 & $+$46:00:01.7& 0.91  0.56     & 2 & B \\
06     \,\ \   02 &  08:13:12.36 & $+$45:59:57.6& 1.07  0.79     & 2 & E \\
07     \,\ \ \ -- &  08:13:15.70 & $+$45:59:57.4& 0.83  0.41     & 2 & E \\
08     \,\ \   03 &  08:13:16.18 & $+$45:59:56.0& 0.96  0.63     & 2 & E \\
09$^{\dagger}$\,\ \ -- &  08:13:15.28 & $+$45:59:54.1& 0.76  0.24     & 2 & E \\
10$^{\dagger}$\,\ \ -- &  08:13:13.56 & $+$45:59:53.4& 0.68  PLR      & 1 & E \\
11      \,\ \   04 &  08:13:12.20 & $+$45:59:53.0& 0.75  0.21     & 2 & B \\
12$^{\dagger}$\,\ \ -- &  08:13:16.74 & $+$45:59:52.9& 1.06  0.78     & 2 & E \\
13      \,\ \   05 &  08:13:14.36 & $+$45:59:52.0& 1.31  1.09     & 1 & E \\
14      \,\ \   06 &  08:13:11.72 & $+$45:59:50.7& 1.36  1.15     & 2 & S \\
15      \,\ \   08 &  08:13:13.85 & $+$45:59:50.7& 0.96  0.63     & 2 & E \\
16      \,\ \   07 &  08:13:14.77 & $+$45:59:50.6& 1.72  1.56     & 1 & S \\
17      \,\ \ \ -- &  08:13:09.23 & $+$45:59:49.5& 1.03  0.74     & 2 & E \\
18      \,\ \   09 &  08:13:13.08 & $+$45:59:48.2& 0.93  0.59     & 2 & S \\
19      \,\ \   10 &  08:13:15.02 & $+$45:59:45.3& 1.56  1.38     & 1 & S \\
20      \,\ \   11 &  08:13:15.33 & $+$45:59:45.2& 2.29  2.17     & 1 & S \\
21      \,\ \   12 &  08:13:15.78 & $+$45:59:44.8& 0.88  0.51     & 1 & E \\
22$^{\dagger}$\,\ \ 13 &  08:13:13.52 & $+$45:59:44.2& 1.02  0.72     & 2 & E \\
23      \,\ \   14 &  08:13:10.94 & $+$45:59:43.0& 0.97  0.65     & 2 & E \\
24      \,\ \ \ -- &  08:13:08.52 & $+$45:59:41.4& 1.19  0.95     & 2 & E \\
25      \,\ \ \ -- &  08:13:15.64 & $+$45:59:41.2& 1.39  1.19     & 2 & N \\
26      \,\ \   15 &  08:13:13.13 & $+$45:59:40.6& 2.11  1.98     & 2 & S \\
27      \,\ \   16 &  08:13:12.94 & $+$45:59:38.4& 1.32  1.11     & 1 & S \\
28$^{\dagger}$\,\ \ 17 &  08:13:13.64 & $+$45:59:36.9& 0.84  0.43     & 2 & E \\
29      \,\ \   18 &  08:13:15.51 & $+$45:59:35.6& 0.85  0.45     & 2 & S \\
30      \,\ \   19 &  08:13:15.92 & $+$45:59:35.6& 0.93  0.59     & 2 & E \\
31      \,\ \   20 &  08:13:12.87 & $+$45:59:35.1& 0.65  PLR      & 2 & N \\
32      \,\ \   21 &  08:13:13.41 & $+$45:59:34.3& 0.69  PLR      & 2 & S \\
33      \,\ \   22 &  08:13:16.11 & $+$45:59:33.9& 0.96  0.63     & 1 & E \\
34      \,\ \ \ -- &  08:13:08.99 & $+$45:59:33.0& 1.32  1.11     & 2 & E \\
35      \,\ \   23 &  08:13:13.69 & $+$45:59:33.0& 1.28  1.06     & 2 & N \\
36      \,\ \   24 &  08:13:13.28 & $+$45:59:32.4& 0.71  PLR      & 2 & N \\
37      \,\ \   26 &  08:13:15.88 & $+$45:59:29.3& 1.90  1.76     & 1 & S \\
38      \,\ \   27 &  08:13:16.23 & $+$45:59:29.2& 0.94  0.60     & 2 & N \\
39      \,\ \   28 &  08:13:17.76 & $+$45:59:29.0& 1.20  0.96     & 2 & B \\
40      \,\ \   25 &  08:13:12.76 & $+$45:59:28.8& 1.62  1.45     & 2 & S \\
41      \,\ \   29 &  08:13:14.04 & $+$45:59:27.3& 1.07  0.79     & 2 & S \\
42      \,\ \   31 &  08:13:12.98 & $+$45:59:26.5& 1.29  1.07     & 1 & S \\
43      \,\ \   32 &  08:13:16.77 & $+$45:59:25.9& 1.12  0.86     & 1 & E \\
44      \,\ \   33 &  08:13:12.63 & $+$45:59:24.6& 0.79  0.33     & 2 & N \\
45      \,\ \ \ -- &  08:13:14.57 & $+$45:59:23.2& 1.15  0.90     & 2 & S \\
46      \,\ \   34 &  08:13:12.37 & $+$45:59:22.3& 1.03  0.74     & 2 & N \\
47      \,\ \   36 &  08:13:16.76 & $+$45:59:21.2& 1.22  0.98     & 1 & S \\
48      \,\ \   35 &  08:13:12.69 & $+$45:59:21.2& 1.18  0.93     & 1 & E \\
49      \,\ \ \ -- &  08:13:15.36 & $+$45:59:18.8& 0.73  0.12     & 2 & N \\
50      \,\ \   37 &  08:13:11.19 & $+$45:59:18.8& 1.42  1.22     & 2 & E \\
51      \,\ \   38 &  08:13:12.72 & $+$45:59:17.3& 1.28  1.06     & 2 & N \\
52      \,\ \   39 &  08:13:12.97 & $+$45:59:17.2& 1.32  1.11     & 2 & S \\
53$^{\dagger}$\,\ \ -- &  08:13:13.64 & $+$45:59:17.1& 0.86  0.47     & 2 & E \\
54      \,\ \   40 &  08:13:14.91 & $+$45:59:15.9& 2.01  1.88     & 1 & S \\
55$^{\dagger}$\,\ \ -- &  08:13:16.12 & $+$45:59:15.8& 0.69  PLR      & 2 & E \\
56      \,\ \ \ -- &  08:13:17.35 & $+$45:59:14.6& 0.81  0.37     & 2 & E \\
\hline
\end{tabular}                                                                    
\label{positions}
\end{table}

\begin{table}
\addtocounter{table}{-1}
\caption[]{(cont.) Positions and sizes for the \object{Mrk~86} neighborhood regions.}
\begin{tabular}{|lccccc|}
\hline
\# \ \ \#' & RA(2000) & DEC(2000) & $r_{1e}$ \ $r_{1e}$' & $e$ & Cl. \\
\hline
(1)      \     (2)&       (3)    &       (4)     & (5)\ \ (6)     &(7)&(8)\\ 
\hline	      
57$^{\dagger}$\,\ \ -- &  08:13:16.05 & $+$45:59:13.1& 0.82  0.39     & 2 & E \\
58      \,\ \   41 &  08:13:16.84 & $+$45:59:12.3& 1.00  0.69     & 2 & E \\
59      \,\ \   44 &  08:13:15.25 & $+$45:59:11.6 & 0.86  0.47     & 2 & S \\
60      \,\ \   43 &  08:13:18.27 & $+$45:59:11.4 & 0.86  0.47     & 2 & E \\
61      \,\ \   42 &  08:13:14.39 & $+$45:59:11.4 & 0.77  0.27     & 3 & F \\
62      \,\ \ \ -- &  08:13:16.04 & $+$45:59:09.8 & 1.28  1.06     & 1 & S \\
63      \,\ \ \ -- &  08:13:12.48 & $+$45:59:09.4 & 0.56  PLR      & 2 & N \\
64      \,\ \   45 &  08:13:16.74 & $+$45:59:08.7 & 1.30  1.08     & 1 & E \\
65      \,\ \ \ -- &  08:13:17.60 & $+$45:59:07.6 & 1.22  0.98     & 1 & E \\
66      \,\ \   46 &  08:13:13.05 & $+$45:59:07.0 & 1.47  1.28     & 2 & S \\
67      \,\ \   47 &  08:13:15.94 & $+$45:59:06.9 & 0.80  0.35     & 3 & F \\
68      \,\ \ \ -- &  08:13:13.75 & $+$45:59:05.9 & 0.62  PLR      & 2 & E \\
69      \,\ \   48 &  08:13:11.94 & $+$45:59:05.2 & 0.95  0.62     & 2 & B \\
70      \,\ \   49 &  08:13:14.28 & $+$45:59:02.8 & 1.43  1.24     & 2 & S \\
71      \,\ \ \ -- &  08:13:15.81 & $+$45:59:02.6 & 0.75  0.21     & 2 & N \\
72      \,\ \   50 &  08:13:15.68 & $+$45:59:02.0 & 1.05  0.76     & 2 & N \\
73      \,\ \ \ -- &  08:13:09.67 & $+$45:59:00.6 & 1.08  0.80     & 2 & B \\
74      \,\ \ \ -- &  08:13:15.52 & $+$45:58:59.0 & 1.08  0.80     & 2 & E \\
75      \,\ \ \ -- &  08:13:15.07 & $+$45:58:53.5 & 0.95  0.62     & 2 & E \\
76$^{\dagger}$\,\ \  -- &  08:13:14.36 & $+$45:58:52.3 & 1.11  0.84     & 1 & E \\
77      \,\ \ \ -- &  08:13:17.80 & $+$45:58:52.1 & 0.87  0.49     & 2 & E \\
78      \,\ \ \ -- &  08:13:12.32 & $+$45:58:48.3 & 1.03  0.74     & 1 & E \\
79$^{\dagger}$\,\ \  -- &  08:13:09.05 & $+$45:58:47.9 & 1.13  0.87     & 2 & E \\
80      \,\ \   51 &  08:13:14.77 & $+$45:58:46.0 & 1.09  0.82     & 2 & E \\
81$^{\dagger}$\,\ \  -- &  08:13:08.64 & $+$45:58:46.0 & 0.64  PLR      & 2 & B \\
82      \,\ \ \ -- &  08:13:14.26 & $+$45:58:45.1 & 0.85  0.45     & 2 & B \\
83      \,\ \ \ -- &  08:13:16.92 & $+$45:58:37.0 & 1.25  1.02     & 2 & B \\
84$^{\dagger}$\,\ \  -- &  08:13:17.39 & $+$45:58:36.6 & 0.91  0.56     & 2 & E \\
85      \,\ \   55 &  08:13:13.69 & $+$45:58:23.4 & 1.37  1.17     & 2 & S \\
\hline                                       
\end{tabular}                                                           
\\                                                                   
\\                                                                  
(1) Knot number (reverse DEC sorted).\\                      
(2) Old knot number as given in GZG.\\
(3) RA(J2000).\\                         
(4) DEC(J2000).\\                        
(5) Radius (in arcsec) at 1\,e-folding.\\                          
(6) Radius (in arcsec) corrected for the atmospheric seeing.\\
(7) Folding (e$^{1}$, e$^{2}$ or e$^{3}$) where colors were measured.\\ 
(8) Classification (see Sect.~\ref{classification}).\\  
PLR= Point-Like Region.\\
$^{\dagger}$ Knot sizes measured on the H$\alpha$ image (see Sect.~\ref{pos+sizes}).
\end{table}

In order to derive the positions and sizes of the regions in the
neighborhood of \object{Mrk~86} we have developed a program called
{\sc cobra} (see Appendix). Briefly, this program selects the image
section where the region of interest is placed. Then, the light
profiles in both image axes are fitted using a straight line, in order
to account for the underlying emission, and two gaussians reproducing
the knot emission profile. The center of the knot is estimated as the
maximum of this latter component. The positions derived for all these
regions are given in Table~\ref{positions} (columns 3 and 4). They
have been numbered in reverse declination order. Most of these regions
were identified on the Johnson $R$-band image, the deepest of those
shown in Fig.~\ref{images} (For $R$=25.6 the signal-to-noise ratio
is 30). However, some regions showed intense H$\alpha$ emission, but
practically no optical or near-infrared continuum emission (see
Table~\ref{positions}). Since more than thirty new objects are
identified in this $R$-band image relative to the Gunn-$r$ image used
in GZG, we have introduced a new notation for the knots. In this
sense, the regions numbered as \#9, \#15,
\#16, \#31 and \#49 in GZG, now are, respectively, \#18, \#26, \#27, \#42 and 
\#70 (see Fig.~\ref{myconts} and Table~\ref{positions}).

The determination of the boundaries of these knots is not an obvious
task. Some authors simply locate the position of the knots by using a
weighted mean of the pixels surrounding a maximum of intensity and
then they use large enough circular apertures to perform the
photometry. This simple approach assume circular geometry for the
knots, a prerequisite which is not always fulfilled. An autonomous
feature, commonly an \ion{H}{ii} region, can be also defined as the
region which is inside its outermost closed contour (Petrosian et
al. \cite{petrosian}) or selecting all pixel that are connected above
a given threshold using predetermined background and noise properties
of the frame (Mazzarella et al. \cite{mazzarella}). All these methods,
rewieved in Fuentes-Masip (\cite{fuentesmasip}), are difficult to
apply in overcrowded fields or when there is a very intense and
variable background emission. We have applied a new method which,
using interactively the {\sc cobra} code, is able to eliminate the
emission of adyacent regions and, specially, the low-frequency
contamination. This contamination is quite relevant in this object due
to the contribution of the underlying stellar population emission to
the surface brightness of the galaxy.

After the underlying emission was subtracted, we estimated the number
of pixels above different thresholds (see Appendix for a complete
description). From the relation between the number of pixels and the
threshold used we determine the physical size of the region at the
e-folding of the equivalent two-dimensional gaussian. The contours
derived in this way will be not circular (see
Fig.~\ref{myconts}). This method avoids sistematic effects
introduced in the contours and sizes determination due to changes in
the underlying emission from one region to another.

The equivalent radii derived are given in Table~\ref{positions}
(column 5). Finally, this procedure allows to take into account the
effect of the Point Spread Function (PSF hereafter) over the sizes
derived. Thus, applying
$\sigma^{2}$\,=\,$\sigma_{\mathrm{measured}}^{2}-\sigma_{\mathrm{seeing}}^{2}$,
we obtain the radii corrected from the PSF effect (see column 6). The
e-folding radius of the PSF in the H$\alpha$ and $R$-band images was
$\sigma_{\mathrm{seeing}}$=0\farcs72. The radii at the e$^{2}$,
e$^{3}$-folding was, respectively, $\sqrt{2}$ and $\sqrt{3}$ times the
e-folding radii derived.

\subsection{Object classification}
\label{classification}

We have spectroscopic data for only 22 of the 85 regions detected in
the neighborhood of \object{Mrk~86} (see Table~\ref{specdata}), 20 of
them in the low resolution spectra and two, \#18 and \#40, in the high
resolution ones. All these regions seem to belong to the galaxy, with
heliocentric emission-lines velocities within the range
400-540\,km\,s$^{-1}$ (see GZG). However, the nature of the remaining
65 objects is much more uncertain.

\begin{table*}
\setcounter{table}{3}
\resizebox{\hsize}{!}{\includegraphics{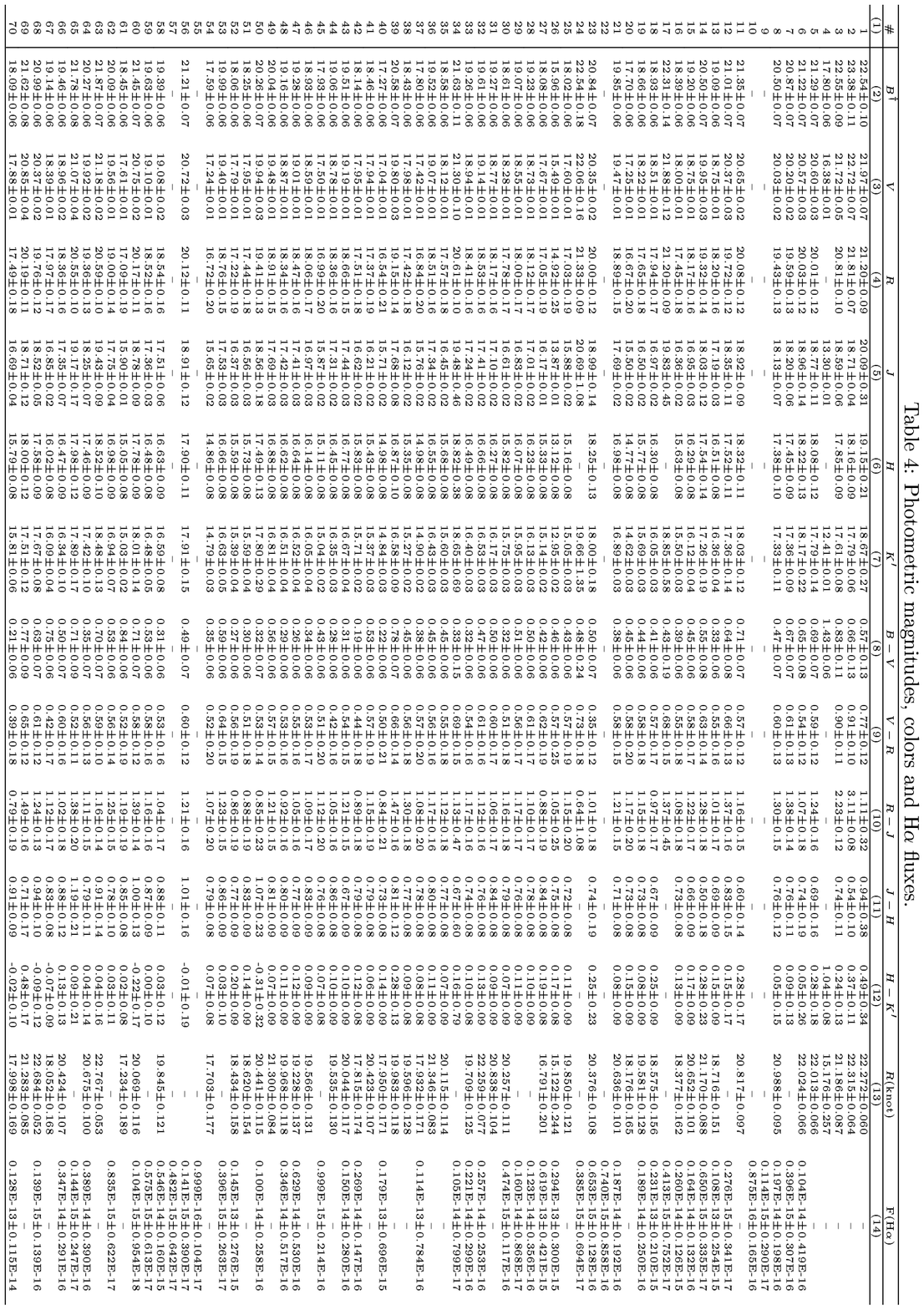}} 
\caption{}
\label{colors} 
\end{table*}

\begin{table*}
\setcounter{table}{4}
\resizebox{\hsize}{!}{\includegraphics{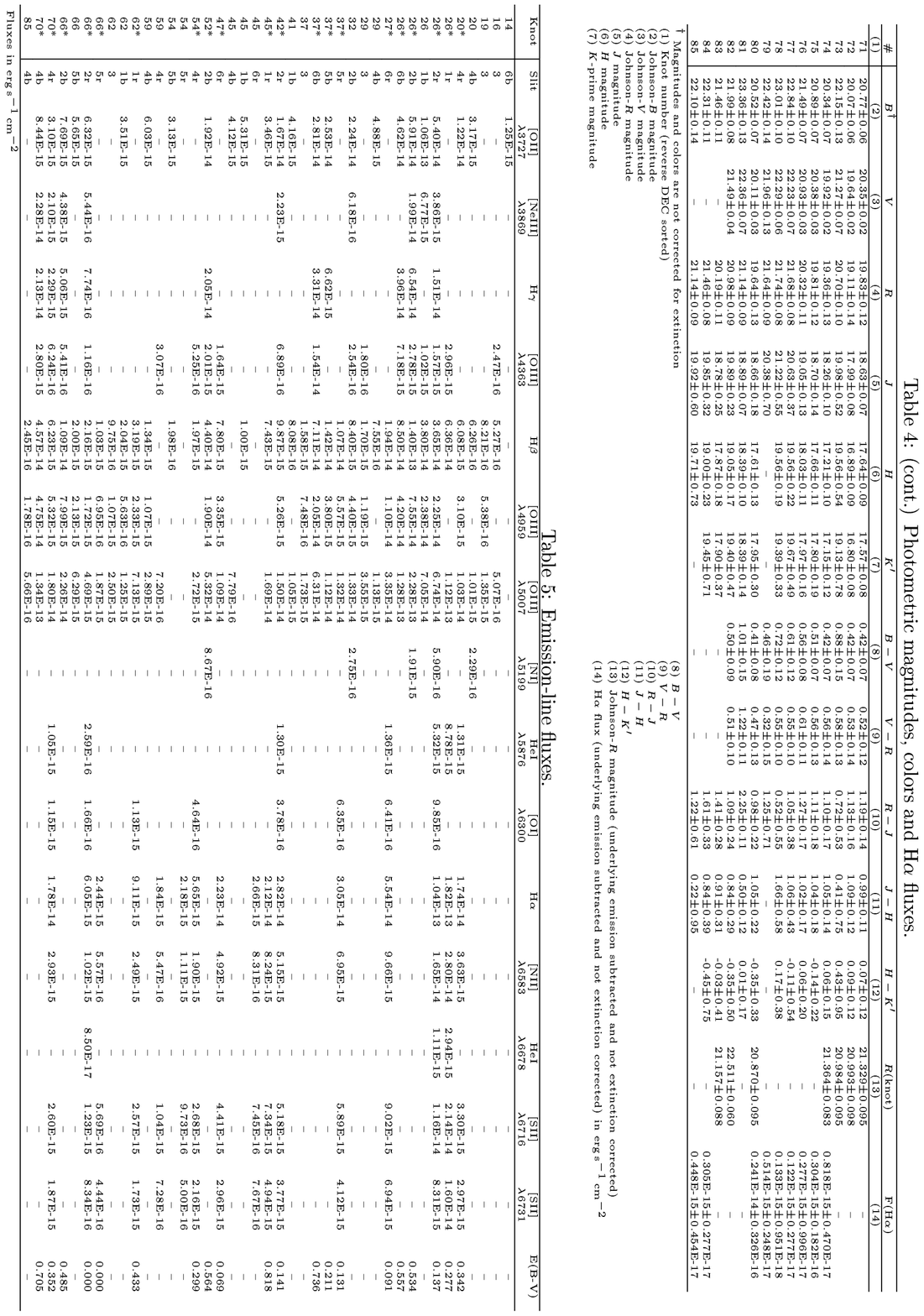}} 
\caption{}
\label{specdata}
\end{table*}

\begin{figure}
\resizebox{\hsize}{!}{\includegraphics{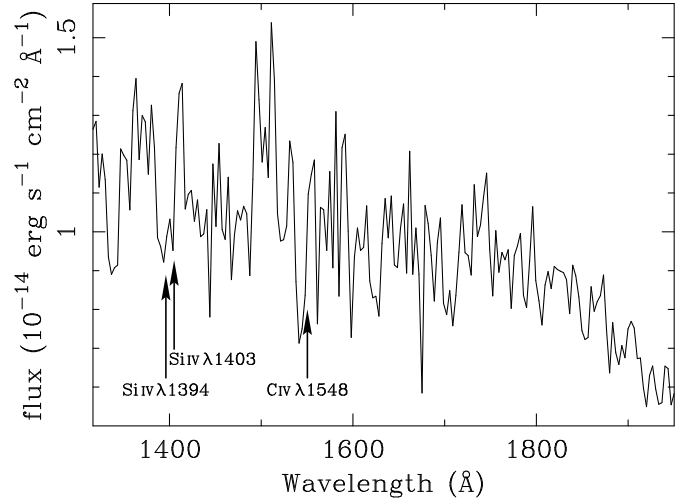}}
\caption{IUE spectrum in the range 1150-1975\AA\ obtained with the SWP camera in the low dispersion mode. The spectrum has been smoothed using a 2 pixels boxcar filter. The \ion{Si}{iv}$\lambda$$\lambda$1394,1403\AA\ and \ion{C}{iv}$\lambda$1548\AA\ spectral lines are shown with a clear P~Cygni profile in the case of the \ion{C}{iv}$\lambda$1548\AA\ line.}
\label{uvspectra}
\end{figure}

\begin{figure*}
\resizebox{\hsize}{!}{\includegraphics{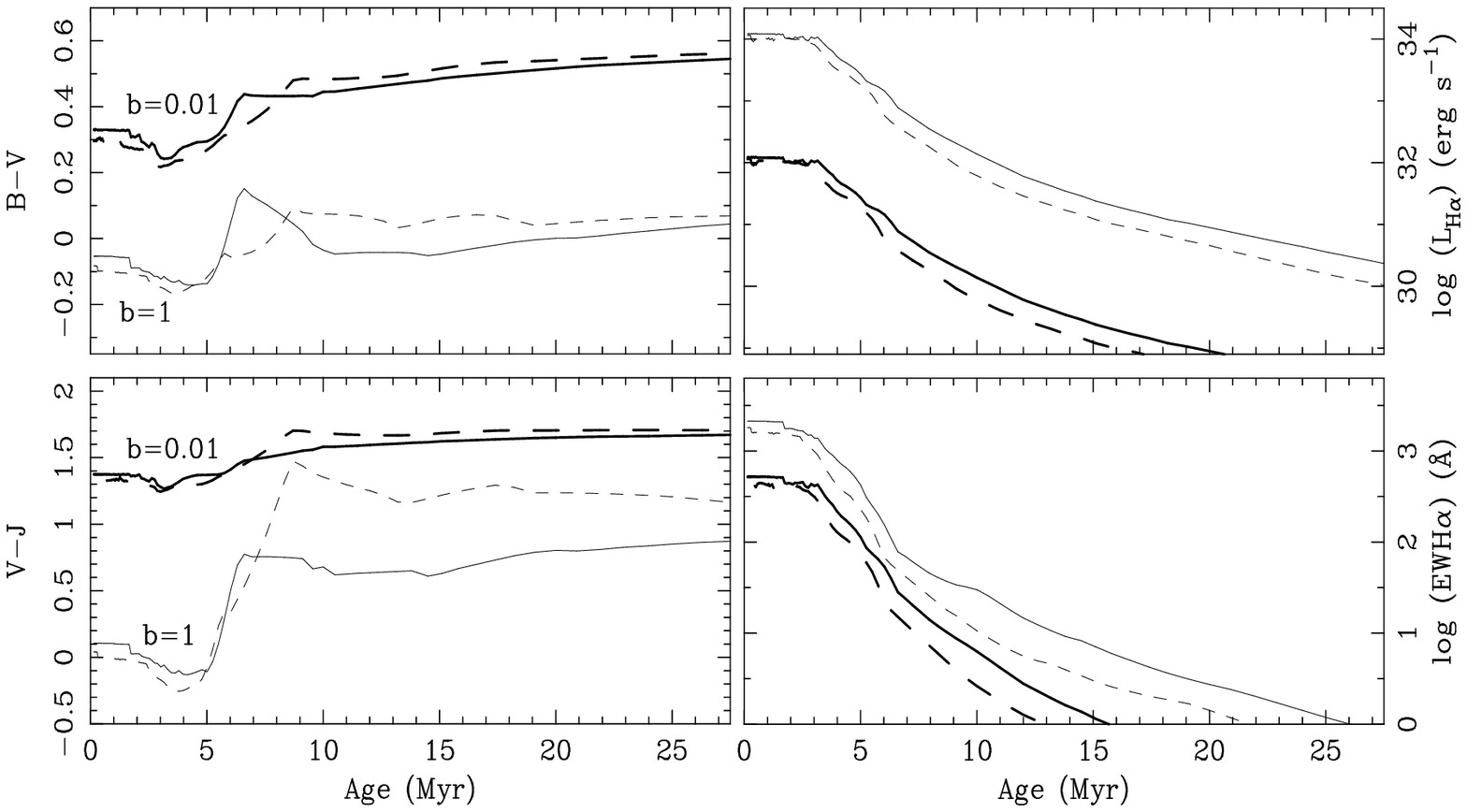}}
\caption{Model predictions. The behavior of the evolutionary synthesis 
models is shown. We give the change in the $B-V$ and $V-J$ colors,
H$\alpha$ luminosity and H$\alpha$ equivalent width for two different
metallicities and burst strengths. $Solid$-$lines$ represent models
with two fifths solar metallicity, and $dashed$-$lines$ represent
solar metallicity models. $Thick$-$lines$ are for $b$=0.01 burst
strength models and $thin$-$lines$ are for pure burst models. These
models have been computed assuming a fraction of 15 per cent of
escaping Lyman photons.}
\label{figmodels}
\end{figure*}

Fortunately, there are many other criteria that provide important
clues about the nature of these objects, {\bf 1)}\,if the host galaxy
and some of these objects show line-emission (e.g., H$\alpha$
emission) within the wavelength range covered by a narrow-band filter
(FWHM$\sim$50-100\AA) they will probably have similar recession
velocities within a range of $\Delta v$$\sim$1000-2000\,km\,s$^{-1}$;
{\bf 2)}\,if one of these objects is placed in a high surface
brightness region of the galaxy, and it has an extended morphology, it
will probably belong to the host galaxy. On the other hand, regions
that do not show line emission and {\bf 3)}\,are placed in galaxy
outer regions, or {\bf 4)}\,show point-like morphology, should not be
classified as belonging to the host galaxy.

In Table~\ref{positions} we mark those regions identified
spectroscopically as belonging to \object{Mrk~86} with an {\bf S}
letter in column 8. Those regions showing photometric H$\alpha$
line-emission are marked with an {\bf E} letter, and with an {\bf N}
those regions with extended morphology placed at short galactocentric
distances. Extended objects placed at large galactocentric distances
have been marked with a {\bf B} (probably background
galaxies). Finally, three point-like sources have been classified as
{\bf F} type (probably foreground stars; \#4, \#61 \& \#67).

It is worthwhile to check whether these point-like objects without
emission lines belong to the galaxy. Their colors (see
Sect.~\ref{starforming}) indicate that objects \#4, \#61 and \#67 are
likely old type stars. Column~13 of Table~\ref{colors} shows the
R-band magnitudes of the regions after being subtracted from the
galaxy background. Using a distance modulus of 29.2 we derive
absolute magnitudes for these stars between 5 and 8 magnitudes too
bright to belong to the galaxy. We conclude that objects \#4, \#61 and
\#67 are effectively foreground stars. In this work we will study only 
the {\bf S}, {\bf E} and {\bf N} type objects.

\subsection{Optical and nIR photometry}
\label{photometry}

Using the contours obtained with {\sc cobra}, we have measured
aperture $BVRJHK$ magnitudes and $B-V$, $V-R$, $R-J$, $J-H$, $J-K$
colors for all the regions in Table~\ref{colors} (columns 2-7 for
magnitudes, and 8-12 for colors). These apertures include both knot
and underlying stellar population emission. Before measuring these
colors we degraded the $BVRHK$ images to the worst seeing $J$-band
image (FWHM$\sim$1.8\arcsec). In addition, we measured integrated
magnitudes in $R$-band for these regions, subtracted from the
underlying emission as determined by {\sc cobra} (Table~\ref{colors};
column 13). Background subtracted H$\alpha$ fluxes were also measured
and they are given in Table~\ref{colors} (column 14). Due to the very
bad seeing of the [\ion{O}{iii}]$\lambda$5007\AA\ image we do not
include photometric [\ion{O}{iii}]$\lambda$5007\AA\ fluxes in our
analysis.

The magnitudes and colors shown in Table~\ref{colors} are measured
quantities and they have not been corrected for extinction.

\subsection{Optical spectroscopy}
\label{spectrares}

The relatively small slit width employed in these observations,
prevents us from obtaining knot total emission line fluxes. We will
derive the emission line ratios in order to characterize the ionized
gas properties in the galaxy star-forming regions. In
Table~\ref{specdata} we give the line intensities measured in a region
of 4\farcs30$\times$2\farcs65 centered in the maximum of the emission
knot section covered by the slit. These line intensities have been
corrected for internal extinction when gas color excesses,
$E(B-V)_{\mathrm{gas}}$, were given. The $E(B-V)_{\mathrm{gas}}$
values have been deduced from the H$\alpha$-H$\beta$ and
H$\beta$-H$\gamma$ Balmer line ratios measured, and assuming the line
ratios predicted for the recombination case-B by Osterbrock
(\cite{osterbrock}). H$\beta$ line intensities have been measured
deblending the absorption and emission components using the
IRAF-STSDAS {\sc ngaussfit} task. The line intensities were measured
separately in both $b$ (blue) and $r$ (red) spectra. When the
line-ratios measured in both arms, typically
[\ion{O}{iii}]$\lambda$5007/H$\beta$, matched for a given region, we
merged both data sets.

\begin{figure}
\begin{center}
\resizebox{7.5cm}{!}{\includegraphics{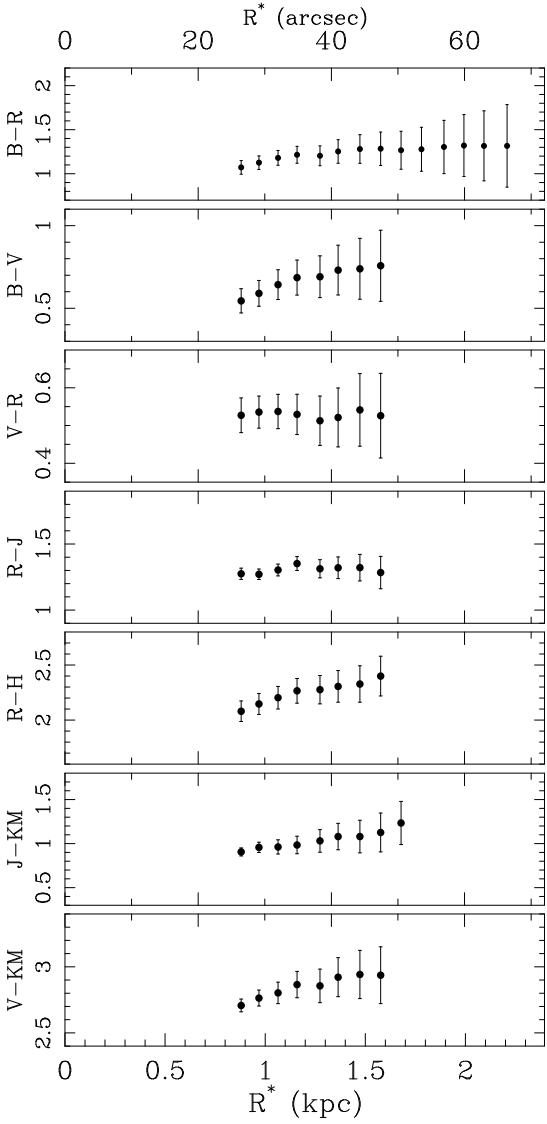}} 
\end{center}
\caption{Color profiles in the outer regions of \object{Mrk~86}.}
\label{colorprofiles}
\end{figure}

\subsection{Ultraviolet spectroscopy}

The aperture used in the UV spectra taken from the IUE Final Archive
(20\arcsec$\times$10\arcsec) was centered in the brightest visual knot
(\#26) with position angle PA=117$\degr$. The UV spectra of this knot
shows the strong absorption lines of
\ion{Si}{iv}$\lambda$$\lambda$1394,1403\AA\ and
\ion{C}{iv}$\lambda$1548\AA\ (see Fig.~\ref{uvspectra}).  Although this
spectrum has been smoothed, the signal-to-noise ratio prevents us from
carrying out a quantitative analysis. However, this spectrum contains
valuable information, showing a clear P~Cygni profile on the
\ion{C}{iv}$\lambda$1548\AA\ line, typical of the fast and dense winds of O-type supergiants.

\section{Evolutionary synthesis models}
\label{models}

In order to derive the physical properties of the different stellar
populations in \object{Mrk~86} we have built a complete set of
evolutionary synthesis models based in those developed by Bruzual \&
Charlot (priv. comm.).

First, we have assumed that the optical-near-infrared spectral energy
distribution (SED) in any region of \object{Mrk~86} can be described
by a young star-forming burst superimposed on an underlying stellar
population. The burst strength parameter, $b$, will describe the mass
ratio between both stellar populations. Then, using this parameter and
the evolution with time of the stellar continuum prediced by the
Bruzual \& Charlot (priv. comm.)  models, we derive the colors and
number of Lyman photons emitted for different composite stellar
populations, with $b$ ranging between 10$^{-4}$ and 1. We have studied
models with metallicities 1/50\,Z$_{\sun}$, 1/5\,Z$_{\sun}$,
2/5\,Z$_{\sun}$, Z$_{\sun}$ and 2\,Z$_{\sun}$, and Scalo
(\cite{scalo}) IMF with $M_{\mathrm{low}}$=0.1\,M$_{\sun}$ and
$M_{\mathrm{up}}$=125\,M$_{\sun}$. The underlying population has been
parametrized using the optical-near-infrared colours measured in the
outer regions of \object{Mrk~86}. We use, $B-V$=0.69, $V-R$=0.52,
$R-J$=1.27, $J-H$=0.99 and $R-K$=2.35, as underlying stellar
population colors, assuming that no significant color gradients are
present (see Fig.~\ref{colorprofiles}). In addition, the mass-to-light
ratio adopted for this stellar population in the $K$-band was
0.87\,M$_{\sun}$/L$_{K,\sun}$ (see Paper~II).

Now, following the procedure described by Gil de Paz et
al. (\cite{nirucm}) and Alonso-Herrero et al. (\cite{aah96}) we
included the contribution to the total flux and colors arising from
the nebular continuum and the most intense emission-lines (i.e.,
[\ion{O}{ii}]$\lambda\lambda$3726,3729\AA, H$\beta$,
[\ion{O}{iii}]$\lambda$4959\AA, [\ion{O}{iii}]$\lambda$5007\AA\ and
H$\alpha$, etc).

We have assumed, in order to compute the nebular continuum emission,
an electron density, $n_{\mathrm{e}}$, of 10$^{2}$\,cm$^{-3}$ and a
temperature, $T_{\mathrm{e}}$, of 10$^{4}$\,K. In addition, from the
analysis of our spectroscopic data (see Paper~II), we adopted a
$N[\ion{He}{ii}]/N[\ion{H}{ii}]$ abundance ratio of 0.12. We have also
assumed that the \ion{He}{iii} abundance is so low that the emission
from recombination to \ion{He}{ii} is negligible.

Finally, we have included the contribution of the emission lines to
the total flux. The contribution of the H$\beta$, H$\alpha$,
Pa$\beta$, Br$_{10}$-Br$_{19}$ and Br$\gamma$ hydrogen emission lines
to the $BVRJHK$ bands was obtained assuming the case-B of
recombination (Osterbrock \cite{osterbrock}) and using the relation
given by Brocklehurst (\cite{brock}). The contribution of the most
intense forbidden lines have been estimated using average
[\ion{O}{ii}]$\lambda\lambda$3726,3729/[\ion{O}{iii}]$\lambda$5007 and
[\ion{O}{iii}]$\lambda$5007/H$\beta$ line ratios, as provided by our
spectroscopic data. Fortunately, the contribution of all the forbidden
lines to the $B$ and $V$ bands is very small. Using the higher and
lower line-ratios measured in the galaxy, this contribution would
range between 1 and 8~per cent for the $B$-band and 2 and 8~per cent
for the $V$-band, for a H$\alpha$ equivalent width (EW hereafter) of
100\AA.

The output of the models will be the optical-near-infrared colors
$B-V$, $V-R$, $V-J$, $J-H$ and $J-K$ of the composite stellar
population, its H$\alpha$ luminosity and equivalent width, and
mass-to-light ratio, parametrized as a function of the burst age,
burst strength and stellar metallicity ($t$,$b$,$Z$).

The Cousins-$R$ magnitudes originally given by the Bruzual \& Charlot
(priv. comm.) models have been converted to Johnson-$R$ magnitudes
using the relation given by Fernie (\cite{fernie}). However, if we
compare the correction predicted by Fernie (\cite{fernie}) in the case
of very red stars ($R_C-R_J$$\simeq$0.25$^{\mathrm{m}}$) with that
measured by Fukugita et al. (\cite{fukugita}) for early-type galaxies,
typically of 0.1$^{\mathrm{m}}$ --with no correction for extinction
applied--, we find differences of about 0.15$^{\mathrm{m}}$. Since the
change in $R_C-R_J$ due to the correction for extinction can not be
higher than 0.02$^{\mathrm{m}}$, this difference has to be attributed
to a difference in the correction between evolved stellar populations
and individual very-red stars. Thus, in the case of the underlying
population analysis, we have applied the mean correction given by
Fukugita et al. (\cite{fukugita}).

\section{Results}
\label{results}

\subsection{Underlying stellar population}
\label{underlying}

The surface brightness profiles given in Fig.~\ref{profiles} show a
clear exponential dominance at regions outer than 
$\sim$1.25\,kpc. We have obtained the parameters of this exponential
component on the deepest $R$-band profile yielding a scale, $\alpha$,
of 0.88$\pm$0.02\,kpc and an extrapolated central surface
brightness, $\mu_{\mathrm{E,0}}$, of
21.50$\pm$0.06\,mag\,arcsec$^{-2}$. Papaderos et al. (\cite{papaI})
obtained quite different values, $\alpha$=0.74$\pm$0.06\,kpc and
$\mu_{\mathrm{E,0}}$=19.42$\pm$0.03\,mag\,arcsec$^{-2}$. This
difference is not surprising considering that the $R$-band profile
given Papaderos et al. (\cite{papaI}) only reaches galactocentric
distances of 2\,kpc.

\begin{figure}
\begin{center}
\resizebox{7.cm}{!}{\includegraphics{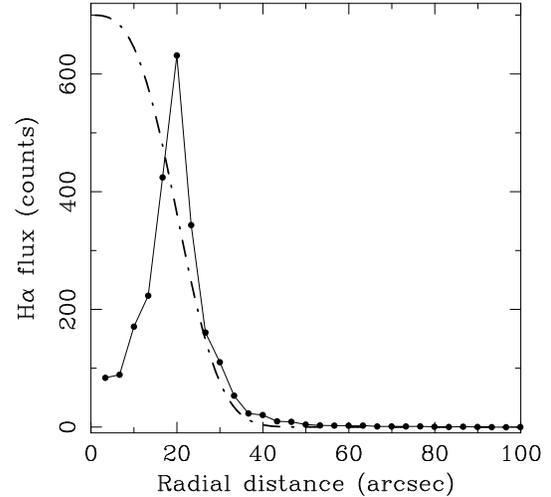}} 
\end{center}
\caption{H$\alpha$ profile obtained using the isophotes measured in the 
$R$-band image ({\it solid-line}). The surface brightness profile of
the {\it plateau} component as parametrized by Papaderos et
al. (\cite{papaI}) in the $B$-band is also shown ({\it
dot-dashed-line}).}
\label{halphaprofile}
\end{figure}

The color profiles in the galaxy outer regions, where the underlying
stellar population dominates the total light profile, are shown in
Fig.~\ref{colorprofiles}. No significant color gradients are
observed at distances larger than 1.2-1.3\,kpc, whereas at shorter
distances a progressive and swallow blueing is derived. This blueing
is probably related with an increment in the light contamination from
high surface brightness star-forming regions associated to the {\it
plateau} component. In Fig.~\ref{halphaprofile} we show the
H$\alpha$ profile compared with the surface brightness profile in the
$B$-band of the {\it plateau} component (see Papaderos et
al. \cite{papaI}). From this figure it is quite clear that the
progresive blueing observed at distances shorter than 40\arcsec\ is
due to currently star-forming regions located in the {\it plateau}
component.

\subsection{Star-forming regions}
\label{starforming}

\begin{figure}
\begin{center}
\resizebox{7.cm}{!}{\includegraphics{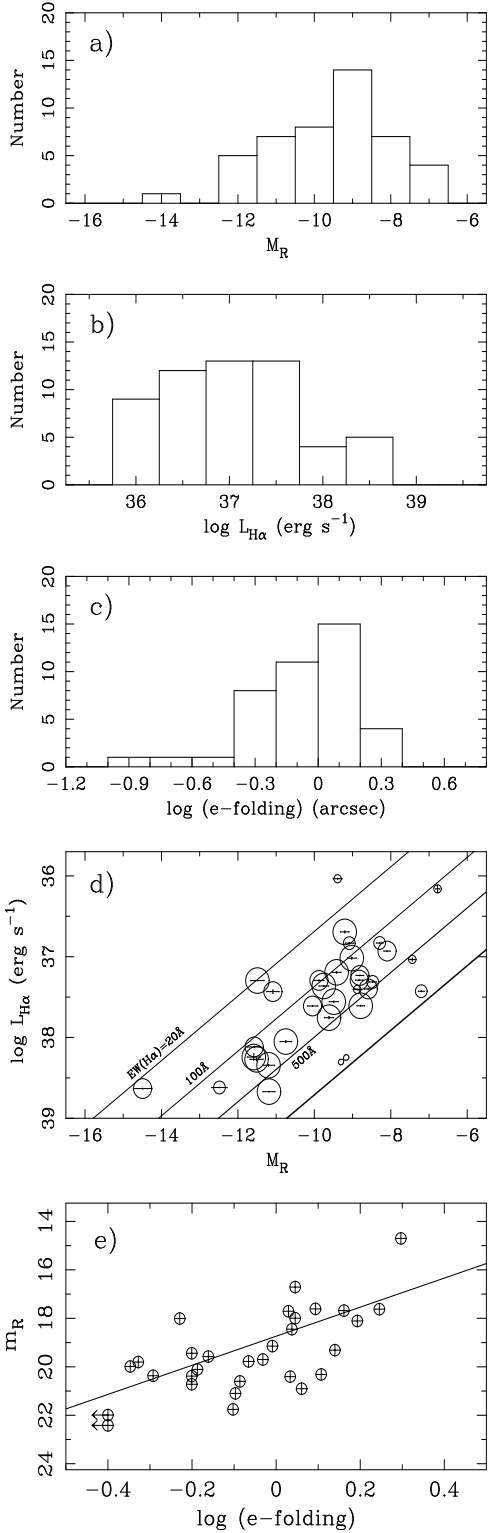}} 
\end{center}
\caption{{\bf a)} $R$-band luminosity distribution for the {\bf S}, 
{\bf E} and {\bf N}-type regions. {\bf b)} H$\alpha$ luminosity
distribution for the galaxy emission-line regions ({\bf S} and {\bf E}
types). {\bf c)} Radius distribution for the {\bf S}, {\bf E} and
{\bf N}-type regions. {\bf d)} H$\alpha$ and continuum $R$-band
luminosities comparison. Lines of constant EW(H$\alpha$) are
drawn. {\bf e)} $R$-band apparent magnitude vs. radius.}
\label{figcolors}
\end{figure}

In Sect.~\ref{classification} we have classified the regions observed
in the \object{Mrk~86} neighborhood as {\bf S}, {\bf E}, {\bf N}, {\bf
F} and {\bf B} objects ({\bf S}, spectroscopically confirmed; {\bf E},
emission-line regions; {\bf N}, extended and diffuse; {\bf F},
foreground stars; {\bf B} background galaxies).

The optical-nIR colors measured for the {\bf F} objects seem to
indicate that the \#4, \#61 and \#67 regions are, respectively, M0-M3,
K2-K5 and G7-K0 spectral type foreground stars. Besides, between the
{\bf B} type objects (background galaxies), we find two very red
objects, \#2 and \#3 regions , with $V-K$ colors of about 5 and
4$^{\mathrm{m}}$, respectively.

Hereafter, we present the results for the study of the {\bf S}, {\bf
E} and {\bf N}-type regions. First, we have analyzed their $R$-band
luminosities in Fig.~\ref{figcolors}a. We observe that this
distribution has a clear maximum at M$_R$=$-$9.5$^{\mathrm{m}}$. This
distribution can not be well fitted using any standard power-law
luminosity function (see, e.g. Elson \& Fall
\cite{elson}, for the \object{LMC} star-clusters LF). 

In Fig.~\ref{figcolors}b we show the H$\alpha$ luminosity
distribution of the galaxy emission-line regions. The H$\alpha$
luminosities used were corrected for internal extinction. In those
cases were spectroscopic data were not available we used a mean color
excess of $E(B-V)_{\mathrm{gas}}$=0.34$^{\mathrm{m}}$. The
distribution obtained is nearly flat in the luminosity range
10$^{37.7}$-10$^{38.7}$\,erg\,s$^{-1}$, with most of the \ion{H}{ii}
regions showing H$\alpha$ luminosities in the range
10$^{36}$-10$^{39}$\,erg\,s$^{-1}$. This luminosity range corresponds
to star formation rates between 6$\times$10$^{-5}$ and
0.06\,M$_{\sun}$\,yr$^{-1}$ (Gil de Paz et al. \cite{nirucm}). The
spatial resolution of the H$\alpha$ image is about 40\,pc. Therefore,
the scarcity of \ion{H}{ii} regions fainter than
10$^{36}$\,erg\,s$^{-1}$ can not be explained as an effect of the
spatial resolution (see Kennicutt et al. \cite{kennicutt}).

It should be noticed that some of these H$\alpha$ emitting regions
showing very faint or null continuum emission (\#7, \#9, \#12, \#50),
seem to be pure gas regions photoionized by distant stellar
clusters. These regions are in many cases associated with the rim of
expanding bubbles (see Martin \cite{martin}, GZG).

\begin{table}
\setcounter{table}{5}
\caption[]{Data for the Mrk~86 starburst component.}
\begin{tabular}{lcc}
\hline
& \multicolumn{2}{c}{Coordinates}\\
\hline
RA(J2000)           & \multicolumn{2}{c}{08$^{h}$13$^{m}$14.69$^{s}$} \\ 
DEC(J2000)          & \multicolumn{2}{c}{$+$45\degr 59\arcmin 21.9\arcsec}  \\
\hline
&    \multicolumn{2}{c}{Photometry}\\
\hline
$B-V$               &    \multicolumn{2}{c}{0.39$\pm$0.06}    \\
$V-R$               &    \multicolumn{2}{c}{0.54$\pm$0.25}    \\
$V-J$               &    \multicolumn{2}{c}{1.59$\pm$0.02}    \\
$J-H$               &    \multicolumn{2}{c}{0.77$\pm$0.08}    \\
$J-K'$              &    \multicolumn{2}{c}{0.85$\pm$0.03}    \\
M$_R$               &    \multicolumn{2}{c}{$-$14.0$\pm$0.2}  \\
Radius              &    \multicolumn{2}{c}{4.3\arcsec}       \\
\hline
&    \multicolumn{2}{c}{Spectroscopic indexes}\\
Slit                &   \#4b & \#1b \\
\hline
D$_{4000}$         &   1.38      &  1.38 \\
Mg2           &   0.06     &  0.04 \\ 
H$\delta$$^{\dagger}$ &   6.0\AA  & 6.2\AA   \\
Fe5270        &   1.20\AA &  --   \\
Fe5406        &   0.74\AA &  --   \\
Region              & 21\farcs45$\times$2\farcs65 & 14\farcs3$\times$2\farcs65\\
\hline
\end{tabular}
\\
$^{\dagger}$ Equivalent width in absorption.\\
\label{starburst}
\end{table}

Now, in Fig.~\ref{figcolors}c we show the radius distribution
measured in the $R$-band image (given by the e-folding radius; see
Appendix). These radii have been corrected for the PSF contribution
(see Sect.~\ref{pos+sizes}). This distribution shows a maximum at
about 1\,arc second, that corresponds to a FWHM of about 55\,pc.

In Fig.~\ref{figcolors}d, the $R$-band and H$\alpha$ luminosities
are compared. The lines drawn suggest that these regions have
equivalent widths of H$\alpha$ ranging between 100-500\AA. Since both,
H$\alpha$ and $R$-band fluxes have been measured subtracted from the
contribution of the underlying population, these EW(H$\alpha$) are
pure \ion{H}{ii} region equivalent widths. In this figure the symbol
size is proportional to the extinction corrected $B-V$ color, using
larger symbols for bluer regions. Finally, we have compared (see
Fig.~\ref{figcolors}e) the knot apparent magnitude and the physical
radius, both measured on the $R$-band image using the {\sc cobra}
program. If all these star-forming regions were optically thin at
these wavelengths and they had similar star densities we would expect
the flux to be proportional to the cube of the radius. In
Fig.~\ref{figcolors}e the best fit to the function $F_R\propto
R^{3}$ is also shown.

\subsection{Central Starburst}
\label{central}

Finally, we have studied the region classified by Papaderos et
al. (\cite{papaI}) as the starburst component (see
Fig.~\ref{myconts}). This component appears as a bright, extended
and not very well defined region in the broad-band $BVRJHK$ images
(see Fig.~\ref{images}).

The integrated blue color measured by Papaderos et al. (\cite{papaI})
in the galaxy central parts, $B-R\simeq$0.9 (0.93$\pm$0.26 in our
work), similar to that observed in Scd and Im galaxy types (Fukugita
et al. \cite{fukugita}), indicates the existence of a young stellar
population superimposed on the evolved underlying component. However,
the absence of significant gas emission (see the H$\alpha$ image in
Fig.~\ref{images}) suggests an intermediate aged dominating
population. In Table~\ref{starburst} we give the colors and
spectroscopic indexes measured for this region (see Trager et
al. \cite{trager98}) at two different slit positions\footnote{The
spectroscopic indexes have been measured using the {\sc reduceme}
package (Cardiel et al. \cite{reduceme}).}. The integrated
optical-near-infrared colors shown in Table~\ref{starburst} have been
measured on the $R$-band image using an aperture with radius
4.3\arcsec\ (1~e-folding), as given by the {\sc cobra} program. This
aperture is marked in Fig.~\ref{myconts} with a {\it thick-lined}
contour. These colors have not been corrected for extinction. The
M$_{R}$ absolute magnitude has been measured subtracted from
underlying emission. The spectroscopic indexes have been measured in
the integrated spectra of the starburst section covered by the slits
\#4b and \#1b. (see Table~\ref{journal}; see also Fig.~1 of GZG).

In Paper~II we will derive the physical properties of this region by
comparing the optical-nIR colors obtained with the predictions of our
evolutionary synthesis models. In addition, the spectroscopic indexes
measured will be compared with those prediced by the Bruzual \&
Charlot (priv. comm.) evolutionary synthesis models.

\section{Conclusions}
\label{conclusions}

In this paper, the first of two, we have presented the observations
and data analysis for the optical-near-infrared study of the Blue
Compact Dwarf galaxy \object{Mrk~86}. We have taken $BVRJHK$,
H$\alpha$ and [\ion{O}{iii}]$\lambda$5007\AA\ images and long-slit
optical spectroscopy for the object. Thus, using all these data,

\begin{itemize}

\item
we have obtained the underlying population surface brightness and
color profiles. The deepest $R$-band image yields an exponential
profile for this component with extrapolated central surface
brigthness of $\mu_{\mathrm{E,0}}$=21.5\,mag\,arcsec$^{-2}$ and scale,
$\alpha$=0.88\,kpc. No significant color gradients are observed in
this component at distances larger 1.2-1.3\,kpc.

\item
We have cataloged and classified all the regions observed in the
neighborhood of the galaxy. If these regions are associated with the
galaxy have been classified as {\bf S} (spectroscopically confirmed),
{\bf E} (confirmed by their photometric H$\alpha$ or
[\ion{O}{iii}]$\lambda$5007\AA\ emission) or {\bf N} (diffuse and
placed in the galaxy central region) type regions. {\bf S} and {\bf E}
type objects are accepted to be galactic \ion{H}{ii} regions. {\bf N}
objects are, probably, evolved stellar clusters, with negligible Lyman
continuum emission. On the other hand, background galaxies and
foreground stars are classified as {\bf B} and {\bf F} types,
respectively.

\item
We have also described the set of evolutionary synthesis models used
in Paper~II. These models are based in those developed by Bruzual \&
Charlot (priv. comm.). They have been obtained for metallicities
between 1/50\,Z$_{\sun}$ and 2\,Z$_{\sun}$, and burst strength ranging
between 10$^{-4}$ and 1. We have included nebular continuum and
recombination and forbidden lines emission. Although the contribution
of the forbidden-line fluxes is quite uncertain in those regions where
spectroscopic data are not available, the contribution of these lines
in the $RJHK$ bands is negligible and lower than 8~per cent for the
$BV$-bands (assuming an EW(H$\alpha$)$\sim$100\AA).

\item The optical-near-infrared colors, fluxes, emission line
intensities of the {\bf S}, {\bf E} and {\bf N} regions have been
measured. Optical and near-infrared colors have been measured. The
physical properties of the galaxy star-forming regions will be derived
in Paper~II, comparing these data with the evolutionary synthesis
models developed. This comparison will be performed using a method
based in the combination of Monte Carlo simulations, a maximum
likelihood estimator, a single linkage clustering anaylisis method and
the Principal Component Analysis (PCA).

\item Finally, we provide optical-near-infrared 
colors and spectroscopic indexes for the central starburst component.

\end{itemize}

\section{Appendix: COBRA program}
\label{appendix}

\begin{figure}
\resizebox{\hsize}{!}{\includegraphics*[130,94][577,361]{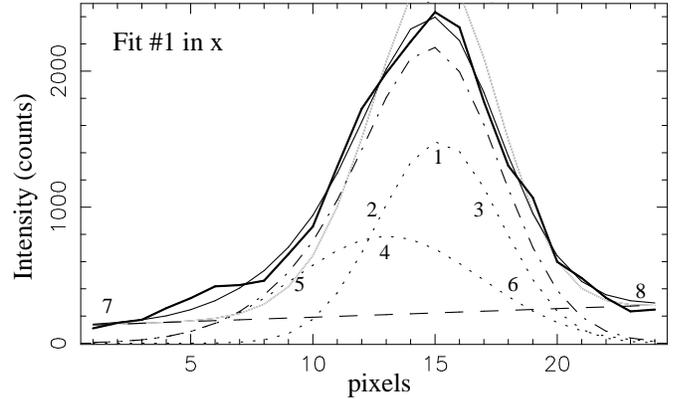}} 
\caption{Example of an observed $x$-axis light profile 
({\it thick solid-line}) and its corresponding best fit ({\it thin
solid-line}). The {\it dotted-lines} correspond to the two gaussian
components, being the {\it dot-dashed-line} the sum of these two
components. The underlying emission (background) is represented by the
{\it dashed-line}. The approximate positions for the 8 points used for
defining the initial set of parameters are also shown. The spatial
scale of the image was 0\farcs333 pixel$^{-1}$.}
\label{perfil}
\end{figure}

Due to the large number of star-forming regions in
\object{Mrk~86} and the intense and variable underlying emission we
developed a program that allowed us to substract this underlying
emission and the occasional contamination from neighbor star-forming
regions. Then, on the underlying emission subtracted image, we
determined the most reliable apertures, sizes, and integrated fluxes
for all these regions. This program was written in {\sc fortran77}
programming language and it was called COBRA {\bf (see {\tt
http://www.ucm.es/info/Astrof/cobra/})}.

COBRA uses different graphic output devices, including {\sc
xwindows} and Postscript, and makes use of the AMOEBA
(Press et al. \cite{press}), PGPLOT, FITSIO and BUTTON (Cardiel et
al. \cite{reduceme}; see also Cardiel \cite{cardiel99}) subroutines.

In this Appendix we will describe the procedure followed to estimate
the underlying emission and the criteria used to determine the sizes
and fluxes of the star-forming regions and stellar clusters in
\object{Mrk~86} using the program COBRA.

\subsection{Light profiles fitting}
\label{a.1}

First, {\bf (1)} a image section including the region of interest
is selected interactively (see the original images in
Fig.~\ref{paneles}). The size of this section necessarily depends on
the spatial scale of the change in the underlying emission and the
proximity of neighbor regions. This section must be chosen in such a
way that the background emission (underlying and neighbor regions
emission) can be reproduced by a line in the $x$ and $y$ light
profiles. Thus, when several relative maxima are placed too close to
match this criterion they were studied as a single region.

After cropping the image section selected, {\bf (2)} the position of
the maximum of the region of interest is marked interactively. Then,
the light profile along the $x$ axis at this position is fitted. 

In order to perform this fit three different components are used, a line
that reproduced the underlying and neighbor regions contamination and
two gaussians that allowed to reproduce the light profile of the region
of interest. In principle, any smooth, monotonic light profile can be
approximated by a serie of Gaussians (somewhat analogous to the
Fourier series), not only the PSF of an image, but also the light
profiles of remote stellar clusters or galactic cores (see
Bendinelli et al. \cite{bendinelli}). Two gaussians are employed in
order to reproduce the light profile of these regions because the use
of a larger number of components led in some cases to solutions with
no physical meaning.

Then, the best fit for the initial light profile using the
minimization subroutine AMOEBA is obtained. The AMOEBA subroutine
needs an initial set of solutions which can be defined by the 8
parameters of the function to minimize (3 for each gaussian and 2 for
the line). These parameters were introduced interactively marking the
approximate position of the center and FWHM for each gaussian and two
extreme values in the profile for the line. In Fig.~\ref{perfil} we
show an example of the positions of the points used to define the set
of initial parameters for the fit. Points 1,2,3 and 4,5,6 led to the
initial parameters for the two gaussians and points 7,8 for the
line. Using this set of parameters, {\bf (3)} the best-fitting light
profile is obtained (see Fig.~\ref{perfil}).

Then, from the output of this fit and introducing the new positions of
the two gaussian maxima, {\bf (4)} all the $x$ axis profiles are
fitted starting from those profiles adjacent to the initial one. The
need of using previous results for the subsequent fits constitutes the main
reason for starting this fitting procedure at the maximum of the
emission region.

Next, {\bf (5)} the same fitting procedure is applied
to the $y$ axis profiles. Finally, after the profiles have been fitted
in both axis, the underlying emission images reconstructed are
averaged and subtracted from the original input image (see several
examples in Fig.~\ref{paneles}). This background-free image is then
used to determine the centers, apertures, sizes and integrated fluxes
of the different regions.

\begin{figure*}
\hspace{2.5cm}\resizebox{!}{0.9\vsize}{\includegraphics{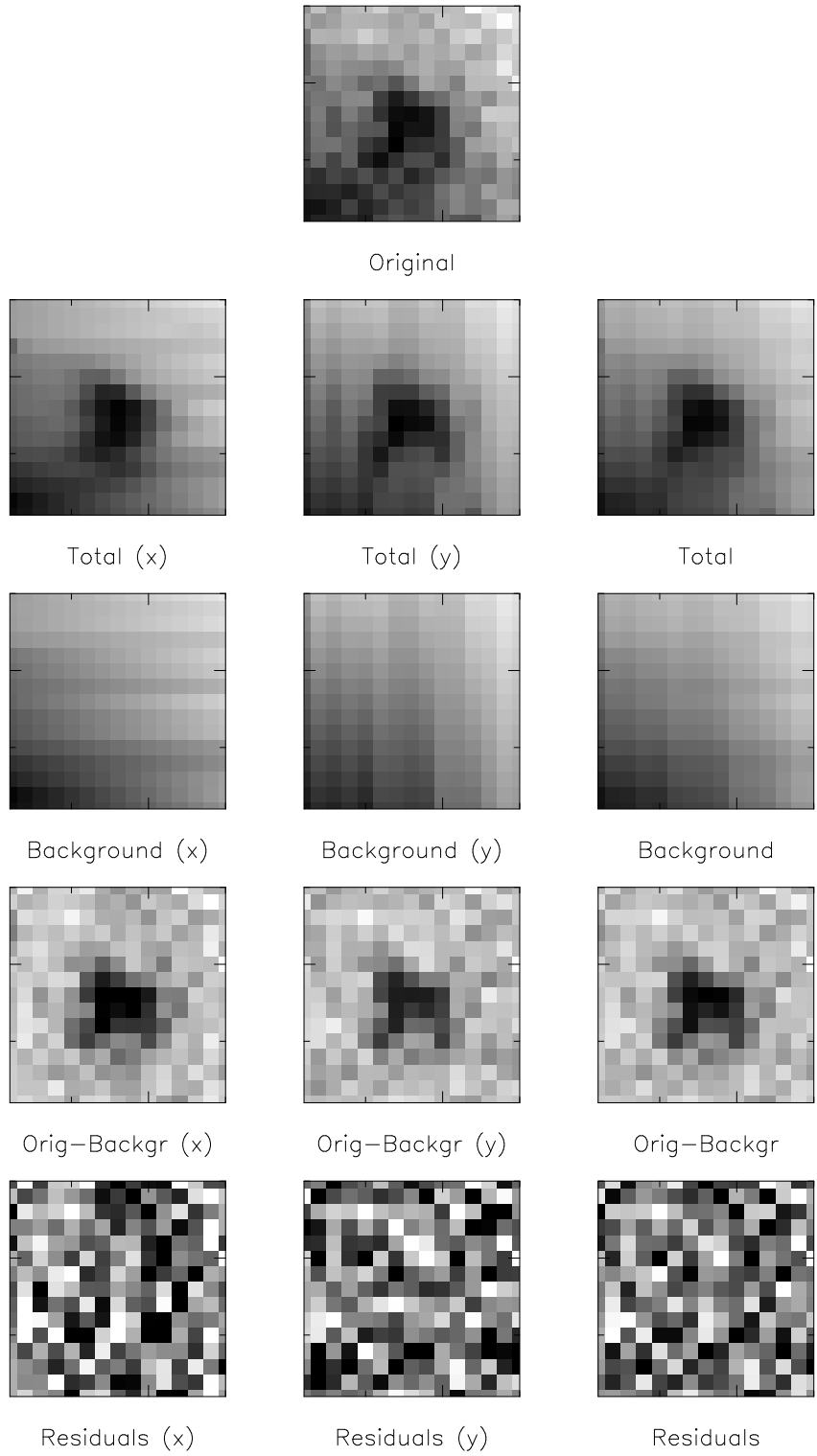}} 
\caption{Original, total (reconstructed), background, original minus 
background model and residuals images in both axis and the average as
obtained for the region \#6 in the $R$-band. The cuts in the original,
total and background images are the same and equal to the maximum and
minimum value in the original image. The cuts in the residuals images
correspond to plus/minus one tenth the difference between the
maximum and minimum in the background averaged image. The residuals
images were obtained as the original image minus the total
(reconstructed) image.}
\end{figure*}
\addtocounter{figure}{-1}

\begin{figure*}
\hspace{2.5cm}\resizebox{!}{0.9\vsize}{\includegraphics{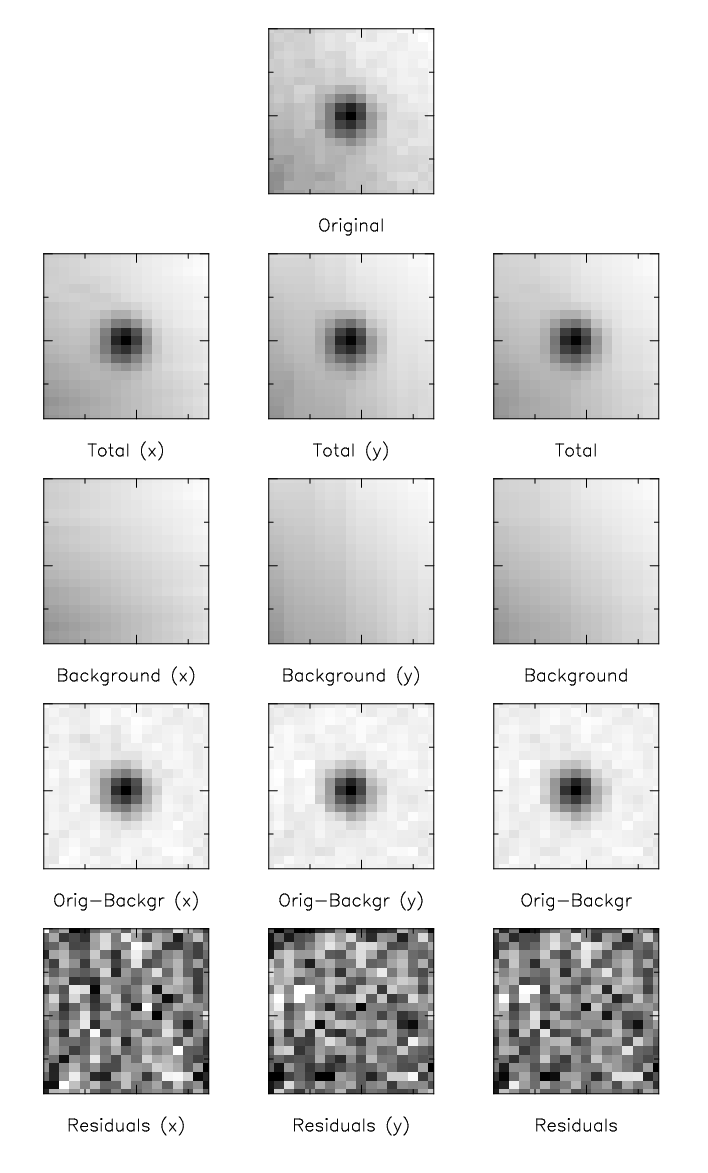}} 
\caption{(cont.) Original, total (reconstructed), background, original minus 
background model and residuals images in both axis and the average as
obtained for the region \#11 in the $R$-band. The cuts in the original,
total and background images are the same and equal to the maximum and
minimum value in the original image. The cuts in the residuals images
correspond to plus/minus one tenth the difference between the
maximum and minimum in the background averaged image. The residuals
images were obtained as the original image minus the total
(reconstructed) image.}
\end{figure*}
\addtocounter{figure}{-1}

\begin{figure*}
\hspace{2.5cm}\resizebox{!}{0.9\vsize}{\includegraphics{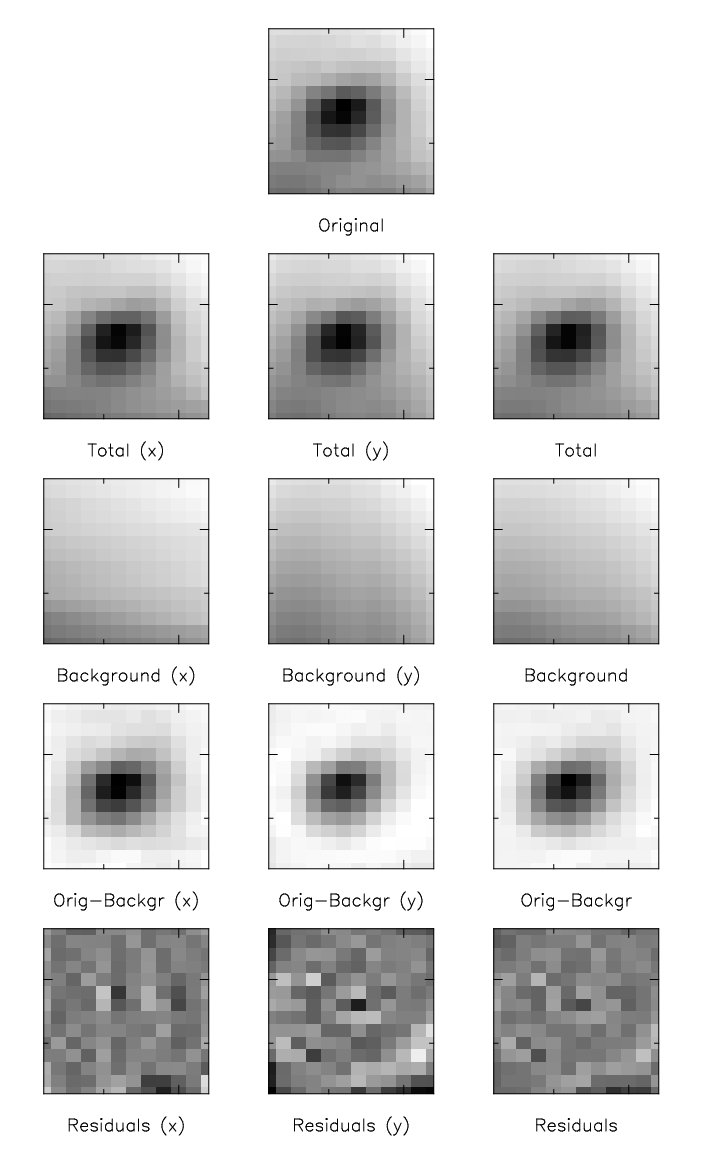}} 
\caption{(cont.) Original, total (reconstructed), background, original minus 
background model and residuals images in both axis and the average as
obtained for the region \#18 in the $R$-band. The cuts in the
original, total and background images are the same and equal to the
maximum and minimum value in the original image. The cuts in the
residuals images correspond to plus/minus one tenth the difference
between the maximum and minimum in the background averaged image. The
residuals images were obtained as the original image minus the total
(reconstructed) image.}
\label{paneles}
\end{figure*}
\addtocounter{figure}{-1}

\begin{figure*}
\hspace{2.5cm}\resizebox{!}{0.9\vsize}{\includegraphics{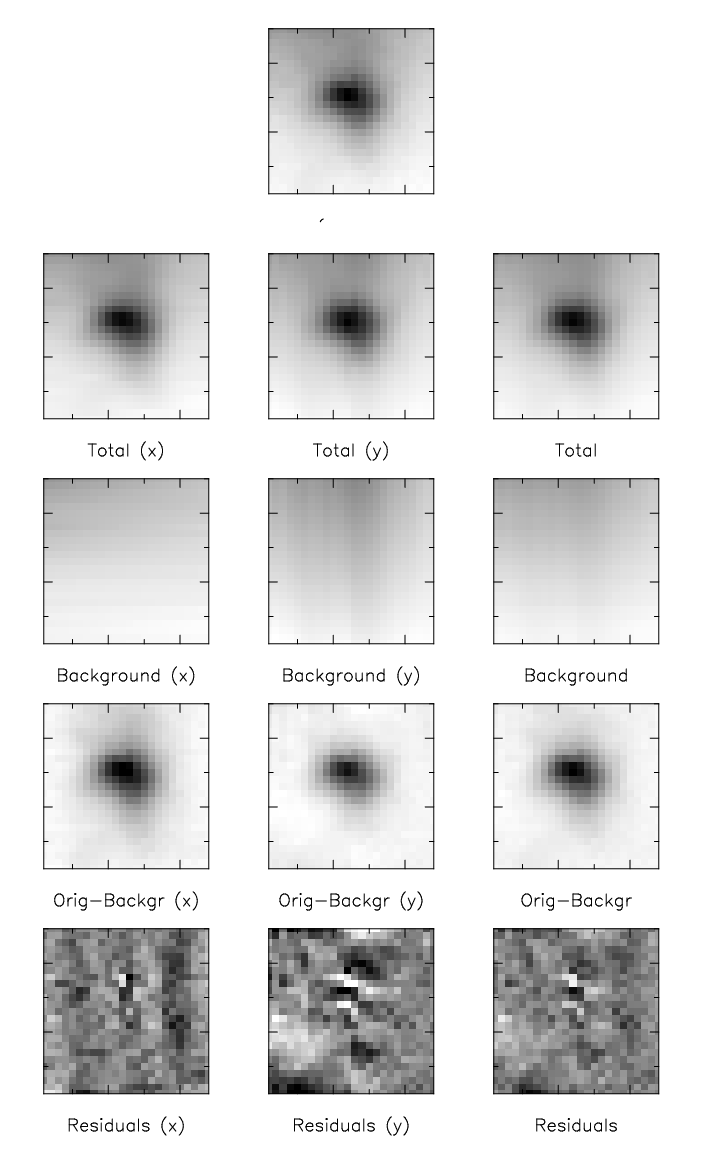}} 
\caption{(cont.) Original, total (reconstructed), background, original minus 
background model and residuals images in both axis and the average as
obtained for the region \#70 in the $R$-band. The cuts in the
original, total and background images are the same and equal to the
maximum and minimum value in the original image. The cuts in the
residuals images correspond to plus/minus one tenth the difference
between the maximum and minimum in the background averaged image. The
residuals images were obtained as the original image minus the total
(reconstructed) image.}
\end{figure*}

\subsection{Position, sizes and fluxes}
\label{a.2}

\begin{figure}
\resizebox{\hsize}{!}{\includegraphics{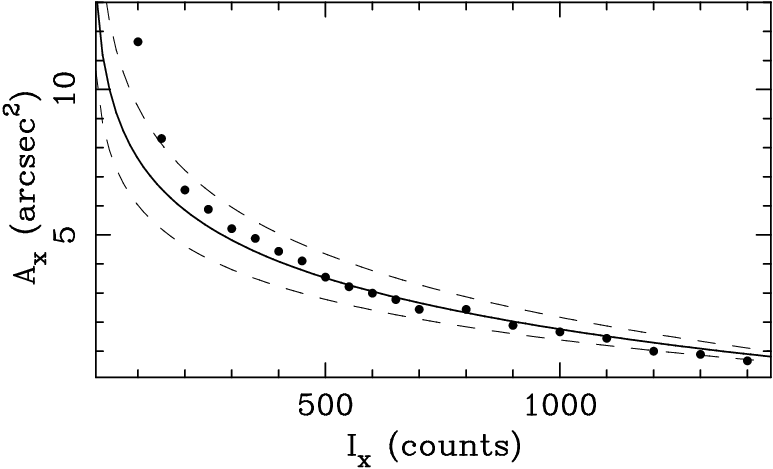}} 
\caption{Change in the area ($A_{\mathrm{x}}$) of the region \#18 
with the threshold, $I_{\mathrm{x}}$. The {\it solid-line} represents
the result expected for a radius of 0\farcs9 at 1 e-folding. The {\it
dashed-lines} correspond to 0\farcs8 and 1\farcs0 e-folding radii.}
\label{axix}
\end{figure}

The knot center is taken as the maximum of the sum of the two gaussian
components. Then, computing the knot surface, $A_{\mathrm{x}}$, above
different thresholds, $I_{\mathrm{x}}$, we can estimate the equivalent
knot e-folding radius, $R_{\mathrm{e-folding}}$, using,
\begin{equation}
R_{\mathrm{e-folding}}=\sqrt{\frac{A_{\mathrm{x}}}{\pi \ln (\frac{I_{\mathrm{max}}}{I_{\mathrm{x}}})}}
\end{equation}
where $I_{\mathrm{max}}$ is the knot maximum intensity. In
Fig.~\ref{axix} we show the change in the area $A_{\mathrm{x}}$ for
different thresholds as measured in the background subtracted image
of the star-forming region \#18 (with 1 e-folding radius of
0\farcs93). The relatively small change deduced for the e-folding
radius with the threshold guarantees the goodness of this size
determination procedure.

The e$^{2}$ and e$^{3}$-folding radii are then computed as $\sqrt{2}$
and $\sqrt{3}$ times the e-folding radius. Then, the knot contours are
derived at these three radii, i.e. at
$I_{\mathrm{max}}$/$I_{\mathrm{x}}$=e,e$^{2}$,e$^{3}$. The knot
total flux is computed as the total flux in the background subtracted
image. In order to ensure that most of the knot flux was included we
derive the knot growing-curve, finding differences not larger than 20
per cent between total and growing-curve extrapolated flux.

\section*{Acknowledgments}
    Based on observations at the Jacobus Kapteyn, Isaac Newton and
    Willian Herschel telescopes operated on the island of La Palma by
    the Royal Greenwich Observatory in the Spanish Observatorio del
    Roque de los Muchachos of the Instituto Astrof\'{\i}sico de
    Canarias and from {\it International Ultraviolet Explorer} archive
    at the ESA VILSPA observatory. Based also on observations
    collected at the German-Spanish Astronomical Center, Calar Alto,
    Spain, operated by the Max-Planck-Institut f\"{u}r Astronomie
    (MPIA), Heidelberg, jointly with the spanish 'Comisi\'on Nacional
    de Astronom\'{\i}a'.
                   
    We are grateful to Carme Jordi and D. Galad\'{\i} for obtaining
    the $V$-band image. We would like to thank C. S\'anchez Contreras
    and L.F. Miranda for obtaining the high resolution spectra,
    N. Cardiel for providing the {\sc reduceme} package and
    C.E. Garc\'{\i}a Dab\'o for the {\sc PlateAstrom} routine. We also
    thank A. Alonso-Herrero for her help in the acquisition and
    reduction of the nIR images. We also acknowledge the anonymous
    referee for several helpful comments. Finally, we thank to
    C.E. Garc\'{\i}a Dab\'o and J. Cenarro for stimulating
    conversations, and M. Sharina for providing a reprint copy of her
    article. This research has been supported in part by the grants
    PB93-456 and PB96-0610 from the Spanish 'Programa Sectorial de
    Promoci\'on del Conocimiento'. A. Gil de Paz acknowledges the
    receipt of a 'Formaci\'on del Profesorado Universitario'
    fellowship from the spanish 'Ministerio de Educaci\'on y Cultura'.

\end{document}